\documentclass[preprint2]{aastex6}

\begin{document}


\title{Near- to mid-Infrared Observations of Galaxy Mergers: NGC2782 and NGC7727}


\author{Takashi Onaka, Tomohiko Nakamura\altaffilmark{1},  
Itsuki Sakon, and Ronin Wu\altaffilmark{2}}
\affil{Department of Astronomy, Graduate School of Science, The University of Tokyo,
Bunkyo-ku, Tokyo 113-0033, Japan}
\email{onaka@astron.s.u-tokyo.ac.jp}

\author{Ryou Ohsawa}
\affil{Institute of Astronomy, Graduate School of Science, The University of Tokyo, Mitaka
Tokyo 181-0015, Japan}

\author{Hidehiro Kaneda}
\affil{Graduate School of Science, Nagoya University, Chikusa-ku, Nagoya 464-8602, Japan}

\author{Vianney Lebouteiller}
\affil{Laboratoire AIM - CEA Saclay, Bat. 709, Piece 267, Orme des Merisiers, 91191 Gif-sur-Yvette,
Fraince}

\and
\author{Thomas L. Roellig}
\affil{NASA Ames Research Center, MS 245-6, Moffett Field, California 94035-1000, U.S.A.}

\altaffiltext{1}{Present address: Recruit Communications Co. Ltd, Tokyo, Japan}
\altaffiltext{2}{Present address: LERMA, Observatoire de Paris, PSL Research University, CNRS,
Sorbonne Universit\'e, UPMC Paris 06, 92190, Meudon, France}

\begin{abstract}
We present the results of near- to mid-infrared (NIR to MIR) imaging and NIR spectroscopic
observations of two galaxy 
mergers, NGC\ 2782 (Arp\ 215) and NGC\ 7727 (Arp\ 222), with the Infrared
Camera on board {\it AKARI}.  NGC\ 2782 shows extended MIR emission
in the eastern side of the galaxy, which corresponds to the eastern tidal tail seen in the
\ion{H}{1}\,21\,cm map,
while NGC\ 7727 shows extended MIR emission in the north of the galaxy, which is
similar to the plumes seen in the residual image at the $K$-band after subtracting a galaxy model.
Both extended structures are thought to have formed associated with their merger events.  
They show excess emission at
7--15\,$\mu$m, which can be attributed to emission from 
polycyclic aromatic hydrocarbons (PAHs), while
the observed spectral energy distributions 
decline longward of 24\,$\mu$m, suggesting
that very small grains (VSGs) 
are deficient.  
These characteristics of the observed MIR spectral energy distribution
may be explained if
PAHs are formed by fragmentation of VSGs during merger events.
The star formation rate is
estimated from the MIR PAH emission in the eastern tail region of NGC\ 2782 and
it is in fair agreement with those estimated from
H$\alpha$ and [\ion{C}{2}]\ 158\,$\mu$m.
MIR observations are efficient for the study of dust processing
and structures formed during merger events.

\end{abstract}




\section{Introduction} \label{sec:intro}
Dust grains retain most of the heavy elements and play significant roles in the thermal balance and chemistry in the interstellar medium (ISM). Thermal emission in the far-infrared (FIR) from submicron-
size dust is often used as a useful measure for the star-formation rate (SFR) for the optically thick case \citep{kennicutt1998},
while the SFR is also known to correlate well with the strengths of the emission features in the mid-infrared \citep[MIR, e.g.,][]{peeters2004, pope2008, shipley2016}, whose 
major bands appear at  3.3, 6.2, 7.7, 8.6, 11.3, and 16.4\,$\mu$m.
They are thought to originate from very 
small carbonaceous dust that contains polycyclic aromatic hydrocarbons  \citep[PAH,][]{tielens2008}
or PAH-like atomic groups, while alternative carriers, such as quenched carbonaceous composite \citep[QCC,][]{sakata1984} or mixed aromatic-aliphatic organic
nanoparticles \citep[MAON,][]{kwok2011}, have also been proposed. 
In the following, we call these emission features PAH features and the carriers PAHs.

The formation and destruction as well as lifecycle of dust grains
in galaxies are vital for the understanding of star-formation and evolution of galaxies.  
The properties and the formation/destruction processes
of carriers of the PAH features are of particular interest 
because they are so conspicuous in the MIR and the relative band ratios should be
useful to study the physical conditions of the environment from which the emission features arise
\citep{draine2001, tielens2008}.
While several theoretical studies have been carried
out \citep[e.g.,][]{dwek1980, jones2011}, little is so far studied observationally about the lifecycle of dust grains in the ISM.   It still remains unclear
where the carriers of the PAH features are formed.  They may be formed in
carbon-rich stars \citep{galliano2008a, paradis2009} or in dense clouds \citep{herbst1991,sandstrom2010}.
Recent observations with{\it AKARI} and {\it SOFIA} suggest that PAHs 
are also formed by fragmentation
of large carbonaceous grains in shocks associated with a supernova, a galactic wind, or an outflow
from a planetary nebula
\citep{onaka2010a, seok2012, lau2016}.  

The PAH features have been detected not only in the disks of galaxies, but also in the halos
of galaxies.  \citet{engelbracht2006} and \citet{beirao2008} reported detection of the PAH
emission in the halo of M82.  \citet{kaneda2010} suggested that PAHs
have been expelled both by the superwind and the galaxy interaction from the disk of M82.  
Survival of PAHs in such harsh environments is another interesting issue for the
study of lifecycle and destruction of PAHs \citep[e.g.,][]{micelotta2010b}.  The PAH features in galaxy halos 
seem to have a lower 7.7\,$\mu$m to 11.3\,$\mu$m band ratio
than that in galactic disks \citep{irwin2006, galliano2008b}. A similar small ratio
is also indicated in the interarm region \citep{sakon2007} as well as in a filamentary
structure associated with the galactic wind \citep{onaka2010a}.  The small ratio may be attributed
to processing of the band carriers in these tenuous regions or could be related to their
formation process.  Observations of
the PAH features in extended structures in galaxies are thus important for the
study of dust processing.

In this paper, we report the results of near-infrared (NIR) to MIR observations of two galaxy mergers,
NGC\ 2782 (Arp\ 215) and NGC\ 7727 (Arp\ 222), with the Infrared Camera (IRC)
on board {\it AKARI} \citep{onaka2007b} to study the properties of dust grains associated
with extended structures produced by merger events 
and their possible processing in violent conditions.  

NGC2787 is a spiral galaxy
classified as Sa(s) Peculiar \citep{sandage1981} or as SABa(rs) Peculiar by \citet{RC3} 
at a distance of 39.5\,Mpc \citep{knierman2013}.  It is a minor merger with an age of
200--300\,Myr old \citep{smith1994, knierman2013}.  NGC\ 2782
shows long tidal tails extending both in the west and east directions in the \ion{H}{1} map
\citep{smith1994}.  A tidal dwarf galaxy candidate (TDGC) was discovered in the eastern side of the galaxy  \citep{yoshida1994}. 
It harbors a starburst both in the central and circumnuclear regions
\citep{jogee1999}.  Gas motion in the central part of the galaxy is studied in detail by
interferometric CO observations \citep{hunt2008}.
NGC\ 2782 was originally classified as a Seyfert galaxy, but no direct sign of AGN
had been obtained \citep{boer1992} until recent X-ray observations found clear evidence
for the presence of a low-luminosity AGN \citep{tzanavaris2007, bravo-guerrero2017}.
Star-formation in the tidal tails was studied extensively based on observations of CO (J=1-0), 
H$\alpha$, and [\ion{C}{2}]\,158\,$\mu$m 
\citep{smith1999, knierman2013}.

NGC\ 7727 is also a merger at a distance of 25\,Mpc and of an age of 1.3\,Gyr old  
\citep{georgakakis2000}.  It is classified as Sa Peculiar by \citet{sandage1981}
and SAB(s)a Peculiar by \citet{RC3}.
The $K$-band image suggests plumes extending to the north of the galaxy from the nucleus \citep{rothberg2004}.  We present the results of IRC NIR to MIR imaging observations of the two galaxies as well as NIR spectroscopy of the central regions of both galaxies.
Star-formation in interacting galaxies has been extensively studied in the IR
\cite[e.g.,][]{smith2007b, brassington2015}.  In this paper, we concentrate on the dust properties
in extended structures in the two galaxies.

\section{Observations and data reduction} \label{sec:obs}
The {\it AKARI}/IRC imaging observations of NGC\ 2782 and NGC\ 7727 were carried out as part of the
mission program 
``ISM in our Galaxy and nearby galaxies'' \citep[ISMNG,][]{kaneda2009a} in the cryogenic
mission phase.  
The observations were made with the two-filter modes of the IRC. 
$10\arcmin \times10\arcmin$ images at bands N3 (3.2\,$\mu$m), N4 (4.1\,$\mu$m), S7 (7\,$\mu$m), and S11 (11\,$\mu$m) were obtained with the AOT IRC02 a;N in the same pointing
observation, while images at bands L15 (15\,$\mu$m) and L24 (24\,$\mu$m) 
were obtained with AOT IRC02 a;L in a different pointing observation.  
Details of the operations of both AOTs are given in
 \citet{onaka2007b}.
NIR IRC spectroscopy of the two galaxies was carried out in the warm mission phase
of {\it AKARI} with the grism mode (AOT: IRCZ4 b;Ns), which provided long slit spectroscopy of 
a $5\arcsec \times 40\arcsec$ area for 2.5--5.0\,$\mu$m with a spectral resolution of
about 0.03\,$\mu$m \citep{ohyama2007}.  
The performance of the IRC in the warm mission phase is given in \citet{onaka2010b}.
A log of the present observations is summarized in Table~\ref{tab1}.  

\begin{deluxetable*}{clllc}
\tablecaption{IRC observation log\label{tab1}}
\tablecolumns{5}
\tablehead{
\colhead{Source name} &
\colhead{Mode} &
\colhead{Pointing ID} &
\colhead{AOT\tablenotemark{a}} &
\colhead{Observation date}
}
\startdata
NGC\ 2782 & imaging (3.2, 4.1, 7, 11\,$\mu$m) & 1400452.1 & IRC02 a;N & 2006 October 31\\
 & imaging  (15, 24\,$\mu$m) & 1401038.1 & IRC02 a;L &  2007 April 29\\
 & spectroscopy (2.5--5.0\,$\mu$m) & 1420364.1 & IRCZ4 & 2009 April 28\\
 & spectroscopy (2.5--5.0\,$\mu$m)& 1420364.2  & IRCZ4 & 2009 April 28\\
NGC\ 7727 & imaging (3.2, 4.1, 7, 11\,$\mu$m) & 1402685.1 & IRC02 a;N & 2007 June 11\\
 & imaging  (15, 24\,$\mu$m) & 1402686.1 & IRC02 a;L & 2007 June 11\\
 & spectroscopy (2.5--5.0\,$\mu$m)& 1420386.1 & IRCZ4  & 2008 December 11
\enddata
\tablenotetext{a}{See \citet{onaka2007b} and \citet{onaka2009} for details of the AOTs in the
cold and warm mission phases, respectively.}
\end{deluxetable*}

The imaging data were processed by the data reduction toolkit for the IRC imaging
version 20110304.
Then, we corrected for the effects of array anomalies in the NIR images (N3 and N4), 
such as muxbleed and column pulldown, by our own developed software, which subtracts fitted
polynomial functions from the data of the affected rows and columns \citep[cf.,][]{carey2004}.  For the MIR images
(S7, S11, L15, and L24), we applied the Richardson-Lucy method with noise suppression using
the wavelet transform \citep{richardson1972, lucy1974, starck1994} to reduce the noise and
restore the images taking account of the point spread functions (PSFs) of each filter band.
Then, the restored MIR images were convolved with a Gaussian to have the same FWHM 
of 7\arcsec.  Finally we took a median of the emission outside of the galaxy and associated extended structures
as the sky background and subtracted it from the observed image.  The uncertainty in the
estimate of the sky background is included in the photometry error.

NIR spectroscopy was made toward the center of the galaxy.  Due to the
pointing accuracy of {\it AKARI}, the actual positions at which spectroscopic 
observations were made were
slightly off the center (see \S~\ref{sec:results} and Table~\ref{tab2}).  The data were processed by the 
data reduction toolkit for the phase 3 (warm mission phase) IRC spectroscopy version 20160324,
which included the latest wavelength calibration \citep{baba2016}.  
The spectrum was extracted for the brightest part of an area of $5\arcsec \times 8\farcs7$ (6 pixels in the slit direction) in the slit for each galaxy.
The central positions of the spectra are listed in Table~\ref{tab2} and the spectrum-extracted areas are 
indicated by the white rectangles 
in Figures~\ref{fig3} and \ref{fig6}.  For NGC\ 2782, we have taken spectra with 
two pointing observations at
slightly different positions ($\sim 1\arcsec$).  The flux levels are slightly different, but the spectral
shapes are almost the same.  We scale the fainter spectrum (Pointing ID: 1420364.1) to the brighter one (Pointing ID: 1420364.2) and co-add them
together to reduce the noise.  The position in Table~\ref{tab1} refers to the spectrum
extracted from the data of 
1420364.2.  NGC\ 7727 was observed only once.

In addition to the IRC observations, we retrieved {\it Spitzer}/Infrared Spectrograph (IRS)
data of NGC\ 2782 from the Combined Atlas of Sources with {\it Spitzer}/IRS Spectra
\citep[CASSIS;][]{lebouteiller2011, lebouteiller2015}.
NGC\ 2782 (AORKEY: 3856986) was observed with the short-low (SL)
and long-low (LL)
modules that provided spectra between 5.2--37.2\,$\mu$m with a spectral resolving power of 64--128 \citep{houck2004}.  
The spectral extraction was performed with Smart/AdOpt \citep{higdon2004, lebouteiller2010}. 
The AdOpt tool enables the simultaneous extraction of blended sources in the IRS apertures. The sources are either point like or they can be described by the convolution of the PSF profile with an intrinsic source shape (e.g., 2D Gaussian). 
Only the spectrum of the central part of the galaxy has a sufficient signal-to-noise ratio 
and is extracted.  The spatial profile of the extracted spectrum is broader than a point source and
it can be well approximated by a compact and an extended component.  
We estimate that the FWHMs of the compact and the extended components are 4\arcsec and 14\arcsec,
respectively, and extract the spectrum of each component in each module (SL and LL).  
The center positions of each component
are listed in Table~\ref{tab2}.  
NGC\ 7727 has not been observed with the IRS.

\begin{deluxetable*}{clllc}
\tablecaption{Positions of the spectrum extracted \label{tab2}}
\tablecolumns{5}
\tablehead{
\colhead{Source} &
\colhead{Instrument} &
\colhead{R.A. (J2000)} &
\colhead{Decl. (J2000)} &
\colhead{Note} 
}
\startdata
NGC\ 2782   & IRC\tablenotemark{a} & 09$^\mathrm{h}$14$^\mathrm{m}$05$\fs$3 
& 40\degr 06\arcmin 53\farcs4 & Extracted for an area of $5\arcsec \times 8\farcs7$.\\
NGC\ 2782c& IRS  & 09$^\mathrm{h}$14$^\mathrm{m}$05$\fs$3 
& 40\degr 06\arcmin 49\farcs7 & Compact component (FWHM = 4\arcsec) \\
NGC\ 2782e& IRS & 09$^\mathrm{h}$14$^\mathrm{m}$05$\fs$2 
& 40\degr 06\arcmin 49\farcs3 & Extended component (FWHM = 14\arcsec) \\
NGC\ 7727 & IRC & 23$^\mathrm{h}$39$^\mathrm{m}$53$\fs$9 
& -12\degr 17\arcmin 30\farcs5 &  Extracted for an area of $5\arcsec \times 8\farcs7$.\\
\enddata
\tablenotetext{a}{The position of the brighter spectrum extracted (see text).}
\end{deluxetable*}

\section{Results} \label{sec:results}
\subsection{NGC\ 2782}\label{sec:results:NGC2782}

\begin{figure*}[ht!]
\plotone{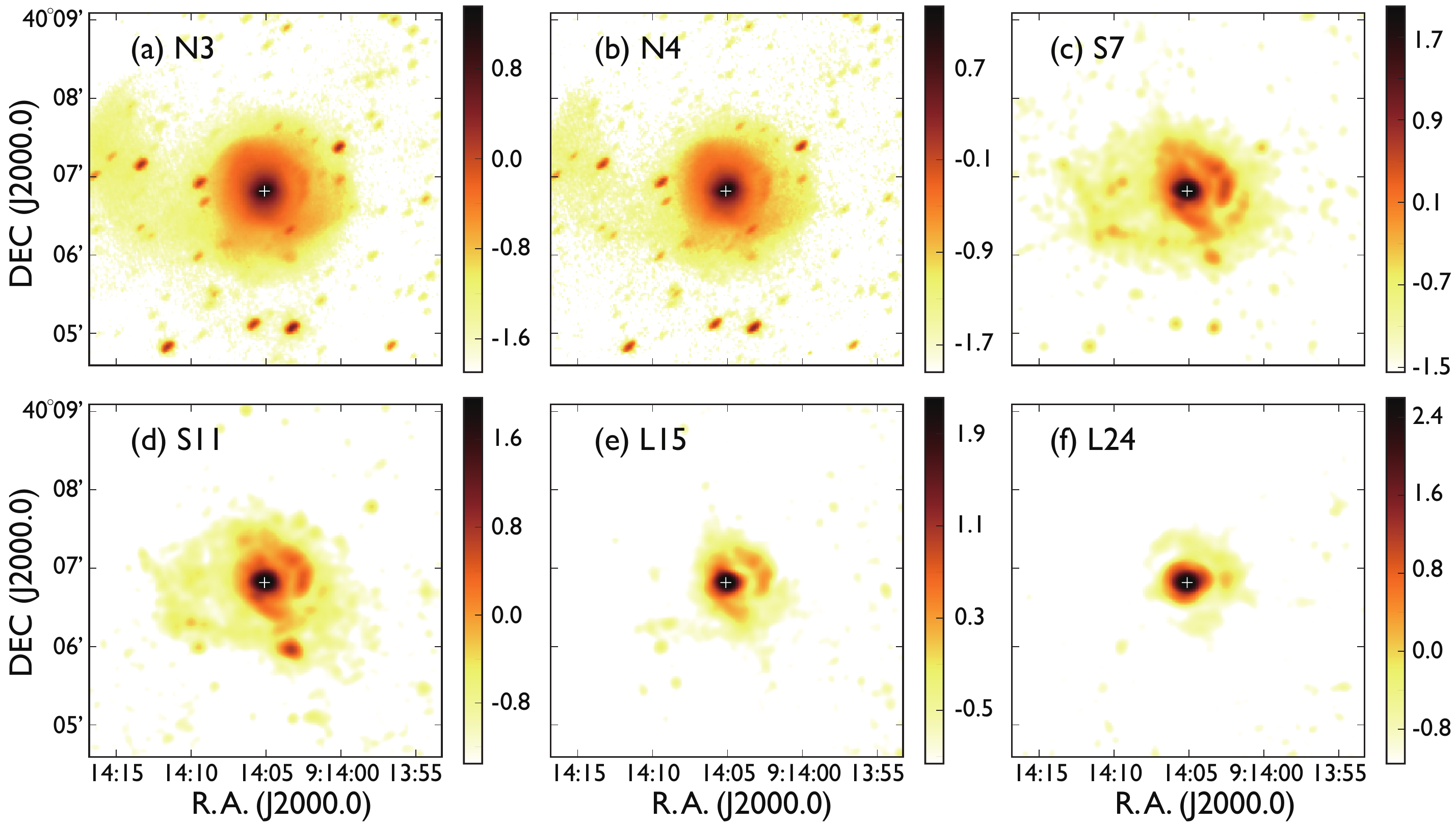}
\caption{IRC images of NGC2782 at the bands (a) N3 (3.2\,$\mu$m), (b) N4 (4.1\,$\mu$m),
(c) S7 (7\,$\mu$m), (d) S11 (11\,$\mu$m), (e) L15 (15\,$\mu$m), and
(f) L24 (24\,$\mu$m).  The center of the galaxy (RA, DEC) = 
(09$^\mathrm{h}$15$^\mathrm{m}$05$\fs$1, 40\degr 06\arcmin 49\farcs0) (J2000.0) is
indicted by the white cross in each image.  The color bar scale is in units of log(MJy sr$^{-1}$).
\label{fig1}}
\end{figure*}

NIR to MIR 6-band IRC images of the central part ($5\arcmin \times 5\arcmin$) of NGC\ 2782 are shown in Figures~\ref{fig1}.  The NIR images of N3 and N4 (Figures~\ref{fig1}a and b) both clearly show extended emission in the eastern side of
the galaxy, which is also seen in optical images \citep{smith1994}. In the MIR images (Figures~\ref{fig1}c--f) it is invisible, suggesting that this extended structure consists only of stellar
components without dust.  In fact, no gas emission has been detected in this region.
It is likely a stellar remnant of the minor galaxy that passed through the major galaxy with its
interstellar materials being stripped off \citep{smith1994}.

The images at the bands S7, S11, and L15 show an arc-like bright emission structure in the western
side of the galaxy (see also Figure~\ref{fig3}), which is also seen as one of the ripples in optical images \citep{smith1994, jogee1999}.  
In the N3 and N4 images, the emission of the stellar disk overlaps with the arc-like ripple structure, 
but it can still be recognized. It becomes faint in the L24 image.  
In addition to the ripple, the S7 and S11 images show extended emission in the eastern side of the galaxy.  We call it the tail structure.  The tail structure is not seen in the N3 and N4 images, suggesting
that it is not associated with 
stellar emission, but comes from the ISM of the galaxy.

\begin{figure*}[ht!]
\plotone{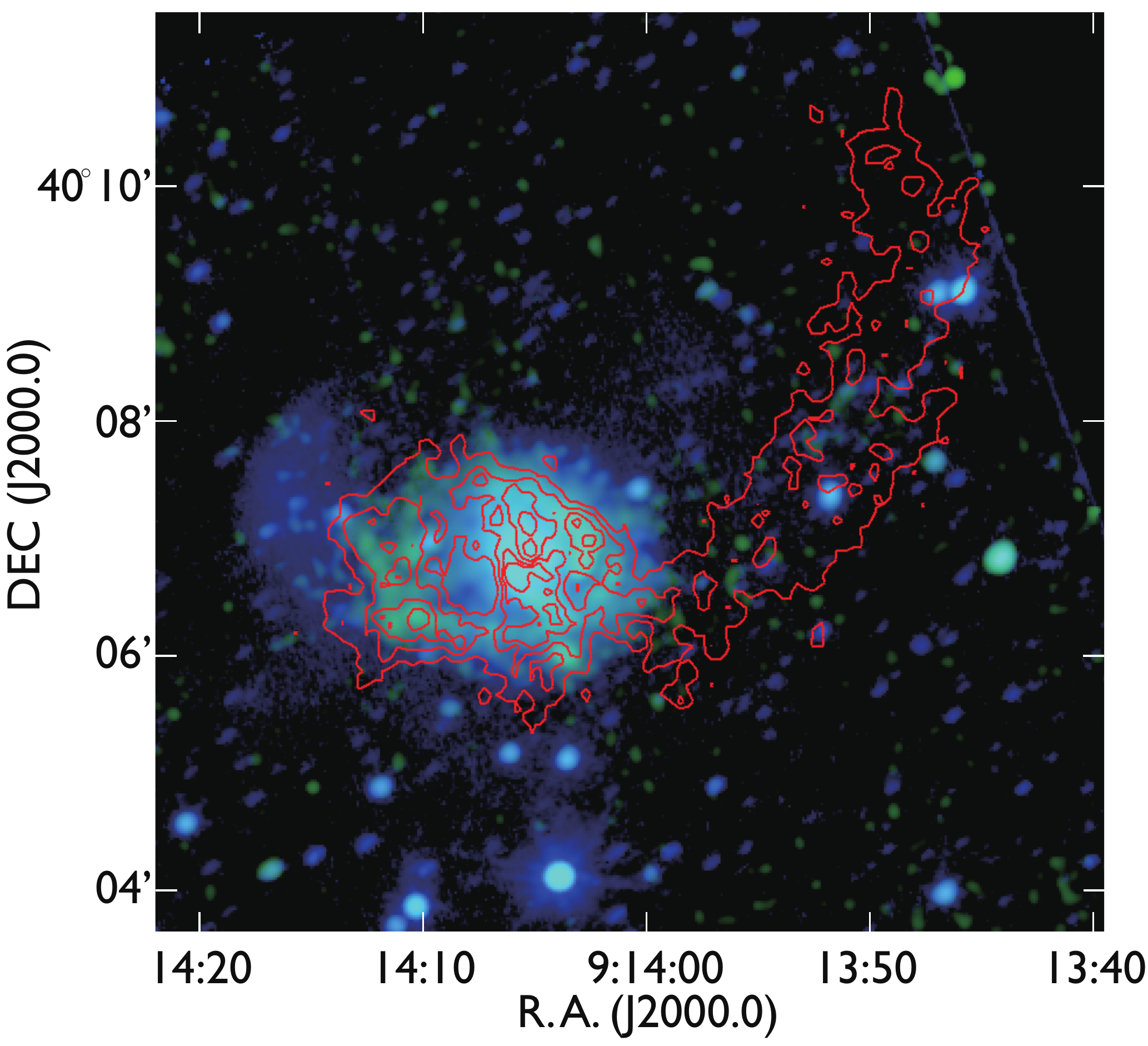}
\caption{Artificial color image of NGC\ 2782.  Blue and green colors correspond to N3 (3.2\,$\mu$m)
and S7 (7\,$\mu$m) intensities, respectively.  The superposed contours in red 
are the VLA \ion{H}{1}\ 21\,cm line intensities
\citep{smith1994} taken from the NASA/IPAC Extragalactic Database (NED). 
\label{fig2}}
\end{figure*}

\begin{figure}[ht!]
  \plotone{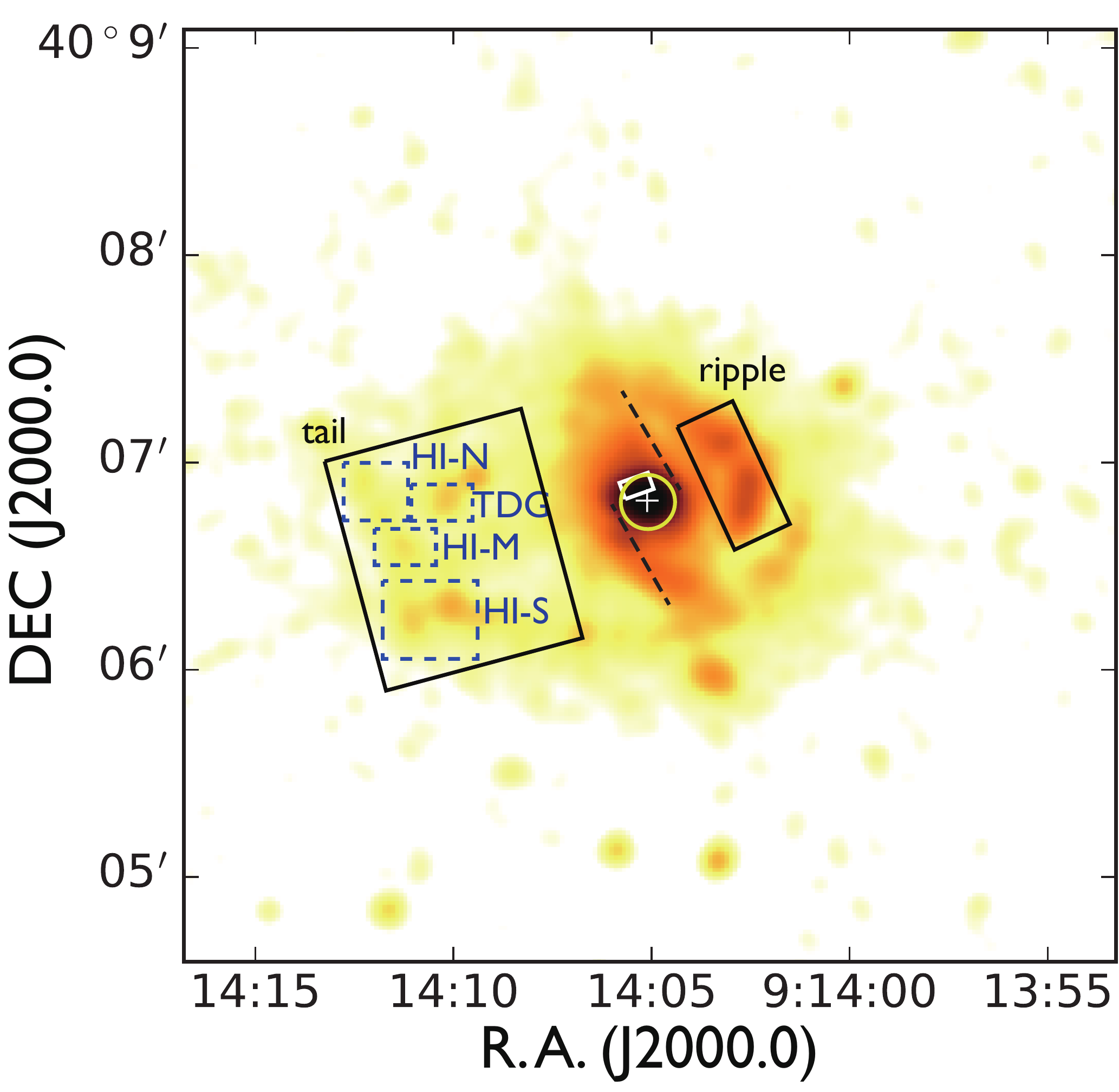}
\caption{Enlarged S7 image of NGC\ 2782.  The white cross symbol indicates the
center position of the galaxy (09$^\mathrm{h}$14$^\mathrm{m}$05$\fs$1, 40\degr 06\arcmin 49\farcs0) and the yellow circle shows the
central 8\arcsec region of the galaxy where the center flux is extracted (see \S~\ref{sec:discussion}).  
The black solid rectangles indicate the areas where the fluxes are extracted for the ripple 
($40\arcsec \times 18\arcsec$) and
the tail ($70\arcsec \times 60\arcsec$)
regions (see \S~\ref{sec:discussion:NGC2782}).
The blue dashed rectangular regions indicate the locations of massive \ion{H}{1} clouds
and the TDGC \citep{knierman2013}.  The black straight dashed lines show the
locations of the dust lanes seen in optical images \citep{jogee1999}.
The white rectangle near the center indicates the area where the IRC NIR spectrum
(Figure~\ref{fig4}a) is extracted.
\label{fig3}}
\end{figure}

To show the tail structure more clearly, an artificial color image of N3 in blue and S7 in green 
was created as shown in Figure~\ref{fig2}, on which the red contours of \ion{H}{1}\ 21\,cm intensity are  superposed \citep{smith1994}.  The \ion{H}{1}\ 21\,cm data show a very long
tail in the western side, at which the IR emission is not detected, and a relatively compact tail
in the eastern side, which delineates the tail structure seen at S7 
(green color) nicely.  
Since both \ion{H}{1} tails must have been formed by the collision event,
the tail structure seen at S7 and S11 is suggested to have also been formed together with
the eastern \ion{H}{1} tail during the collision event.
In the further east, the stellar remnant is clearly seen only at N3 (blue color).
The white color of the center and ripple regions suggests that they are bright both at N3 and S7.

\begin{figure}[ht!]
\plotone{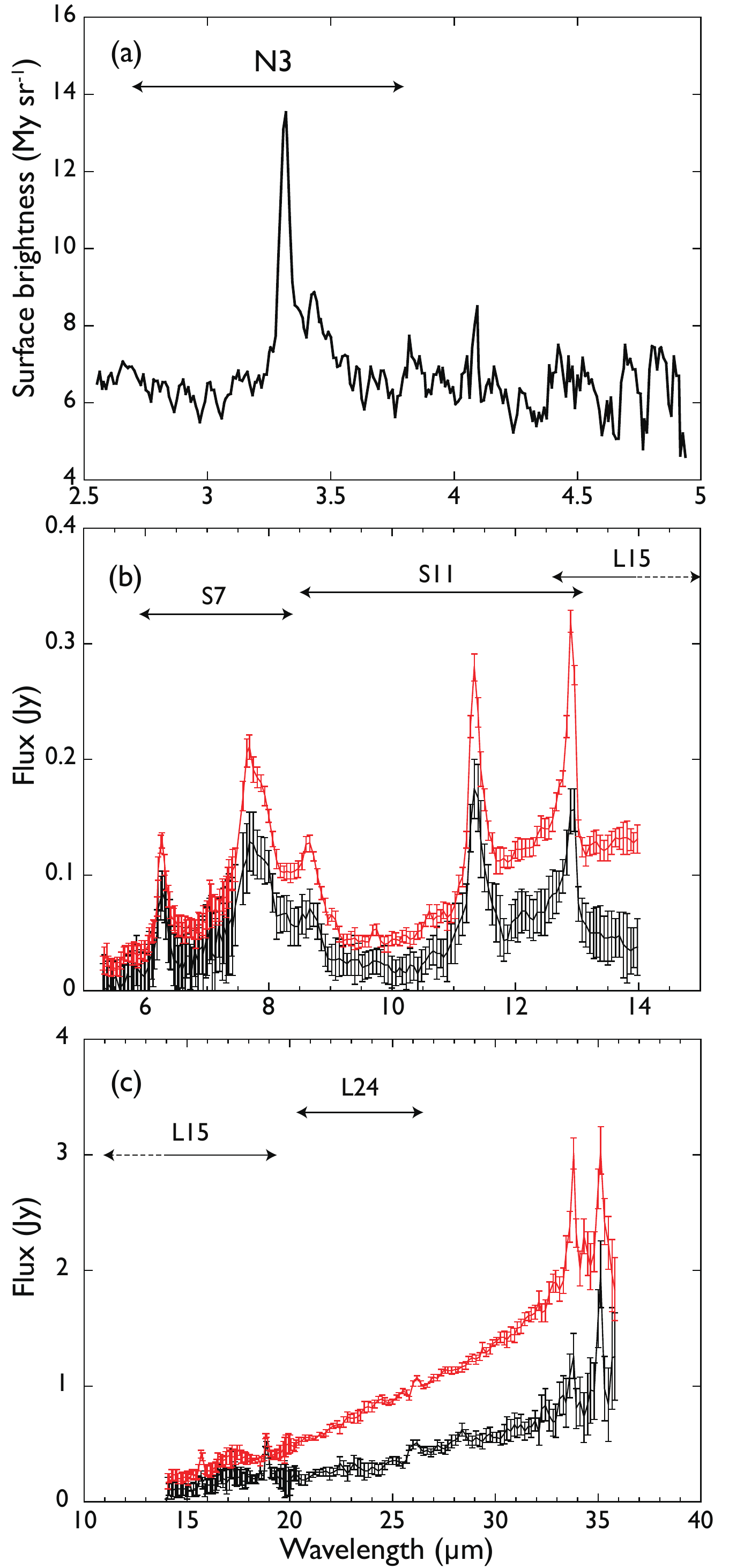}
\caption{Infrared spectra of the center region of NGC\ 2782. (a) {\it AKARI}/IRC NIR spectrum extracted from
the area indicated  by the white rectangle in Figure~\ref{fig3}.  
(b) {\it Spitzer}/IRS SL spectrum and (c) LL spectrum. The IRS spectra are extracted at the center
of the galaxy (Table~\ref{tab2}), assuming the source sizes as described in the text.
In (b) and (c), the central compact component is indicated by the red lines and the
extended component by the black lines.  The arrows in the top of each figure
indicate the e-folding
response range of the IRC filters \citep{onaka2007b}.  The response of L15 filter spans the
SL and LL spectral ranges.
\label{fig4}}
\end{figure}

Figure~\ref{fig3} shows an enlarged S7 image to indicate these structures
clearly.  The blue dashed rectangles in the tail structure 
indicate the locations of massive \ion{H}{1} clouds 
as well as the TDGC \citep{smith1994, knierman2013}.
The emission at S7 shows good correlation with the regions of the massive \ion{H}{1} clouds.  
There may be an enhancement in S7 even at the
position of the TDGC.  Figure~\ref{fig3} also suggests that the emission at S7 becomes
strong along the optical dust lanes \citep{jogee1999} that are indicated by the black straight dashed lines.

Figure~\ref{fig4} shows the NIR and MIR spectra of the central part of NGC\ 2782.
Note that units of the spectra are different; the NIR spectrum (Figure~\ref{fig4}a) is in units of MJy\,sr$^{-1}$,
while the spectra in Figures~\ref{fig4}b and c are in units of Jy, for which the source sizes are assumed
as described in \S\ref{sec:obs}.  
The NIR spectrum (Figure~\ref{fig4}a) is extracted from the white rectangle in Figure~\ref{fig3}
and the MIR spectra are taken nearly at the center of the galaxy (Table~\ref{tab2}).
The observed positions of each spectrum are not exactly the same and thus the relative
comparison of the spectra needs caution.
The NIR spectrum clearly shows the presence of the PAH 3.3\,$\mu$m band as well as the emission from aliphatic C-H bonds
at 3.4\,$\mu$m.  The emission of \ion{H}{1} Br$\alpha$ at 4.05\,$\mu$m is barely
seen, but no other gas emission features are detected.  In the IRS SL spectra, 
the PAH features at 6.2, 7.7, 8.6, and
11.3\,$\mu$m are clearly seen in both compact and extended components
(Figure~\ref{fig4}b).  The [\ion{Ne}{2}]\ 12.8\,$\mu$m overlaps with
the PAH 12.7\,$\mu$m feature.  
The compact component is slightly brighter than the extended component up to around
12\,$\mu$m, above which the flux of the compact component sharply increases.
In the IRS LL spectra, the PAH 17\,$\mu$m complex, 
[\ion{Ne}{3}]\ 15.6\,$\mu$m, [\ion{S}{3}]\ 18.7 and 33.5\,$\mu$m, 
and [\ion{Si}{2}]\ 34.8\,$\mu$m are
seen in both components.  
The compact component is relatively brighter in the ionized gas lines,
while the extended component has relatively brighter [\ion{Si}{2}], suggesting
the dominance of emission from photo-dissociation regions (PDRs).
There seems to be no clear
sign for the presence of an AGN, such as strong [\ion{Ne}{5}]\,14.3\,$\mu$m or [\ion{O}{4}]\,26\,$\mu$m  emission, 
in both spectra \citep{armus2007, farrah2007}.  Upper limits of both line intensities relative to the [\ion{Ne}{2}]\,12.8\,$\mu$m intensity 
are estimated as about 20\%, 
which suggests that the AGN contribution to the MIR emission is less than 10\% \citep{petric2011}.
The presence of an AGN is not confirmed
by the present spectroscopy.  

\subsection{NGC7727}\label{sec:results:NGC7727}
\begin{figure*}[ht!]
\plotone{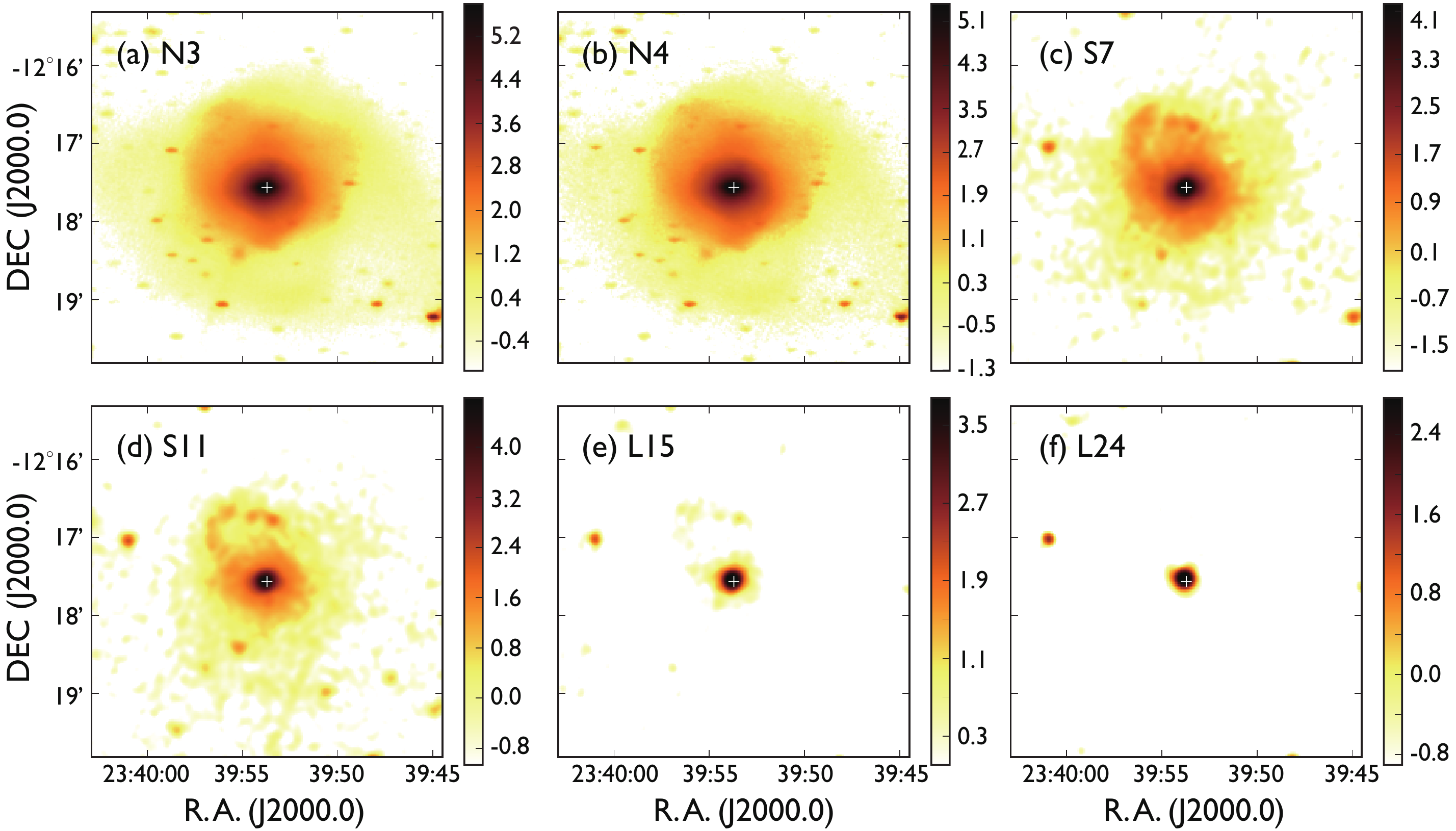}
\caption{IRC images of NGC\ 7727 at the bands (a) N3 (3.2\,$\mu$m), (b) N4 (4.1\,$\mu$m),
(c) S7 (7\,$\mu$m), (d) S11 (11\,$\mu$m), (e) L15 (15\,$\mu$m), and
(f) L24 (24\,$\mu$m).  The center of the galaxy (RA, DEC) = 
(23$^\mathrm{h}$39$^\mathrm{m}$53$\fs$7, -12\degr 17\arcmin 34\farcs0) (J2000.0) is
indicted by the white cross in each image.  The color bar scale is in units of log(MJy sr$^{-1}$).
\label{fig5}}
\end{figure*}

\begin{figure}[ht!]
\plotone{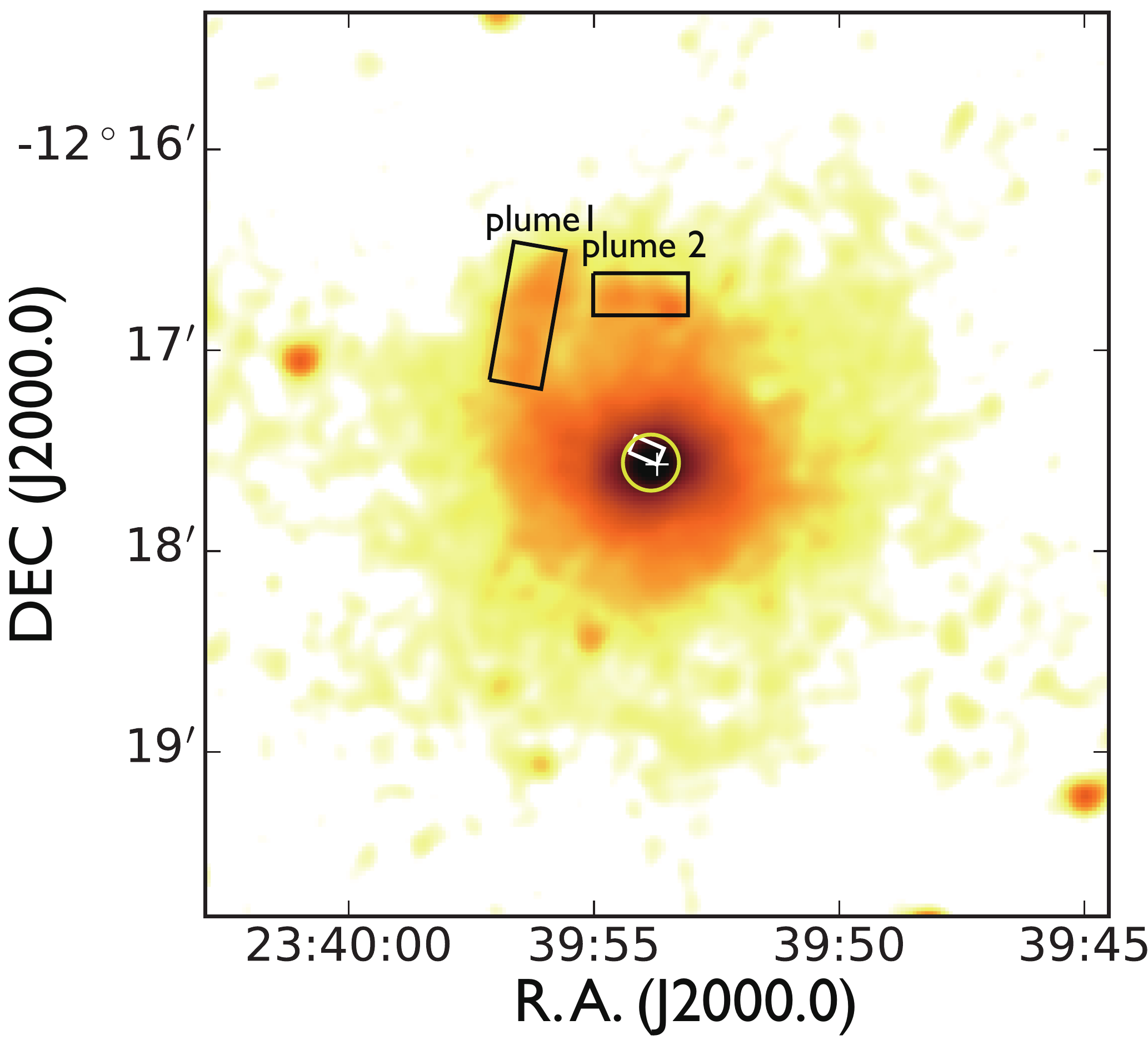}
\caption{Enlarged S7 image of NGC\ 7727.  The white cross indicates the
center position of the galaxy (23$^\mathrm{h}$39$^\mathrm{m}$53$\fs$7, -12\degr 17\arcmin 34\farcs0) and 
the white rectangle near the center shows the area where the IRC NIR spectrum
(Figure~\ref{fig7}) is extracted.
The black rectangles indicate the areas where the fluxes are extracted for
the plume-like structures plume 1 ($40\arcsec \times 15\arcsec$) and plume 2 ($27\arcsec \times 12\arcsec$, see \S\ref{sec:discussion:NGC7727}).
The yellow circle of 8\arcsec in radius shows the region where the flux at the center of the galaxy is
estimated.  Note that the center of the yellow circle is at the center of the
intensity distribution, which is slightly shifted from the
center of the galaxy indicated by the while cross (see text).}
\label{fig6}
\end{figure}

Figure~\ref{fig5} shows 6-band IRC images of the central part ($5\arcmin \times 5\arcmin$) of NGC7727.  Compared to NGC\ 2782, the flux is concentrated on the galaxy center.
In the S7 image, plume-like structures
appear in the northern side of the galaxy 
(see also Figure~\ref{fig6}), which becomes clearer in the S11 image.
In the L15 image, the structures become fainter, but their trace 
is still visible.  They almost fade away at L24.  In the N3 and N4 images, part of the plumes are
barely seen as emission extending to the north-east.
Figure~\ref{fig6} shows an enlarged S7 image to indicate these structures clearly as well
as the region where the IRC NIR spectrum is extracted.  
Similar, but more diffuse plume structures are seen in the residual image in $K$-band after subtracting a galaxy model
\citep{rothberg2004}.
The structures are named
plume 1 and plume 2 as shown in the figure.  We note that the
intensity distribution around the center is slightly shifted toward east from the center
position of the galaxy indicated by the white cross.  We estimate the intensity of the central part of the galaxy
within the yellow circle (see \S\ref{sec:discussion}).

\begin{figure}[!ht]
\plotone{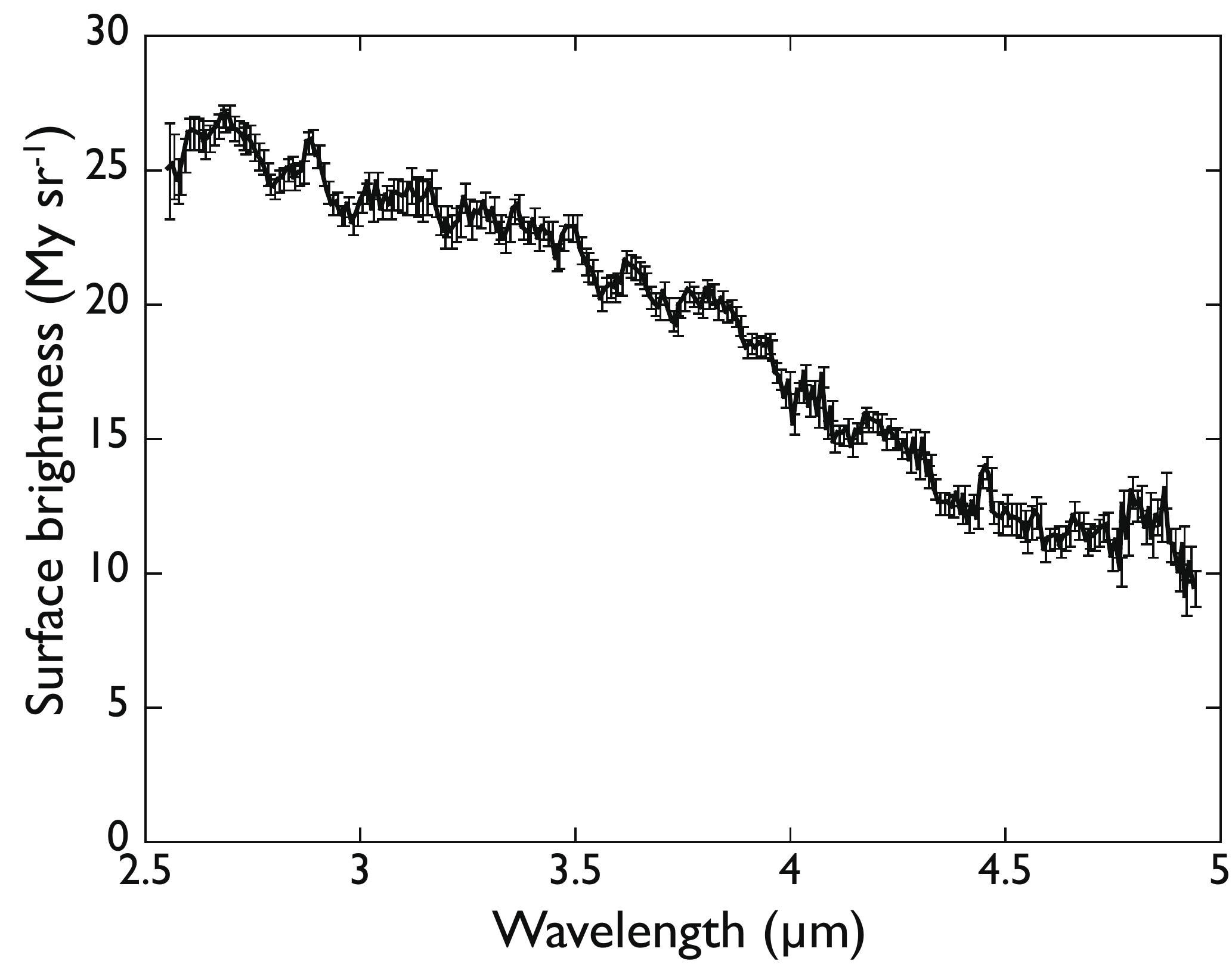}
\caption{NIR spectrum of the center region of NGC\ 7727. The region where the
spectrum was extracted is indicated by the white rectangle in Figure~\ref{fig6}.  
\label{fig7}}
\end{figure}

Figure~\ref{fig7} shows the {\it AKARI}/IRC NIR spectrum of the central region of NGC\ 7727.
There seem to be no clear emission nor absorption features 
in the spectrum.  A shallow, broad dent at around 4.5\,$\mu$m may be attributed to
the CO fundamental vibration absorption, but the present data do not have a sufficient quality to confirm it. 
Compared to the NIR spectrum of NGC\ 2782 (Figure~\ref{fig4}a), the NIR continuum of NGC\ 7727 
is bluer, suggesting the dominance of the contribution from the stellar continuum over the emission
from the ISM.  The absence of \ion{H}{1} recombination lines
and the PAH features suggests low star-formation activities in the central
region of NGC\ 7727.

\section{Discussion} \label{sec:discussion}
\subsection{NGC\ 2782} \label{sec:discussion:NGC2782}
The MIR images at S7 and S11 show a tail structure in the eastern side of the galaxy, 
which has a good
correspondence with the structure of the eastern \ion{H}{1} tidal tail.  In Figure~\ref{fig4}, the spectral
range that each filter of the IRC covers is indicated by the arrows.  
The emission of the PAH 6.2 and 7.7\,$\mu$m bands dominates band S7 \citep{ishihara2007}.
In the S11 image, the 11.3 and 12.7 \,$\mu$m PAH bands are dominant features together with
[\ion{Ne}{2}] 12.8\,$\mu$m.  Although there is no spectroscopic information for this observation
with {\it AKARI}, 
we assume that the S7 and S11 images
trace the PAH emission also for regions
other than the center \citep[cf.,][]{onaka2010a}.
On the other hand, the L15 and L24 do not contain strong dust band features and they trace
mostly the continuum emission.  

\begin{figure*}[htb!]
\plotone{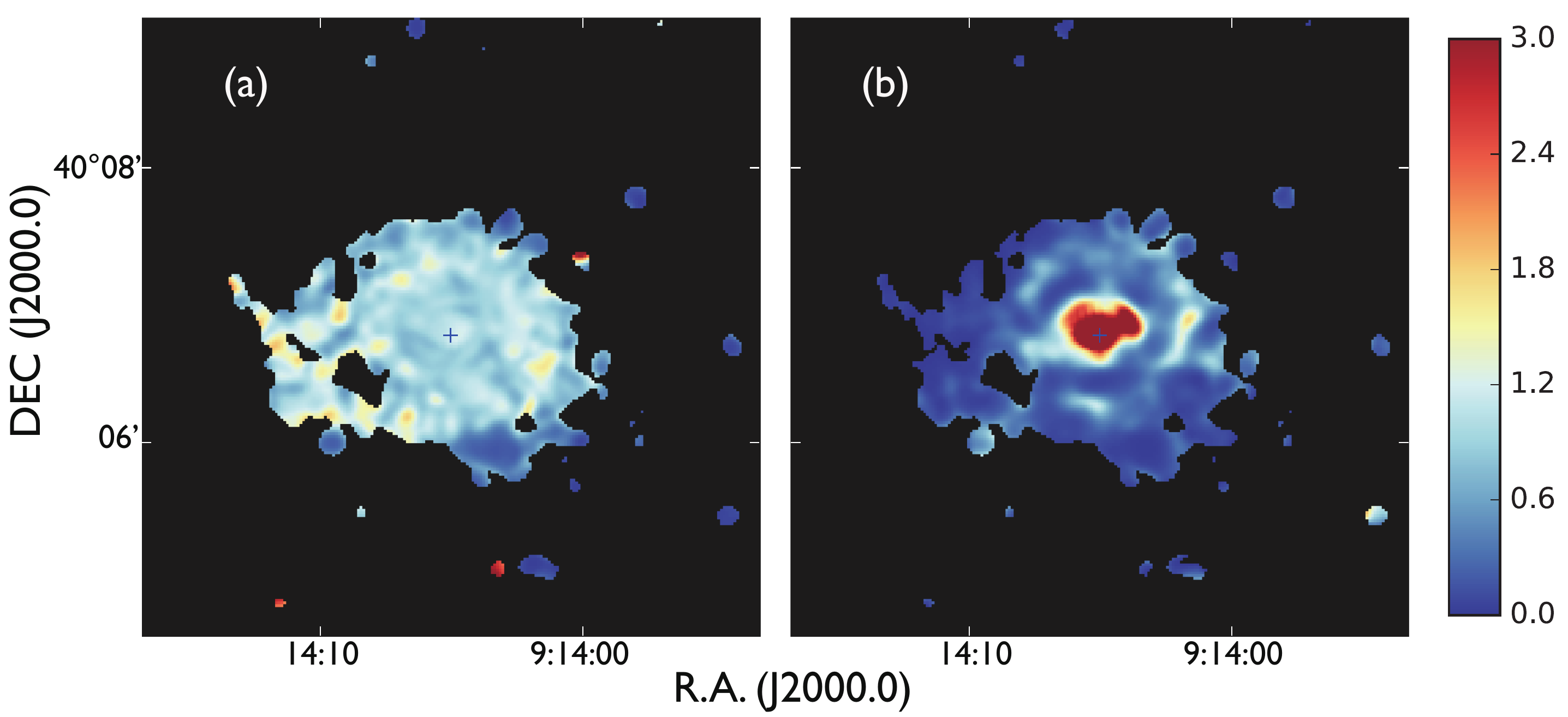}
\caption{Band ratio maps of NGC\ 2782. (a) S7 to S11 and (b) L24 to S11 ratios.  The pixels with fluxes
less than 4$\sigma$ are masked and shown in black.
\label{fig8}}
\end{figure*}

Figure~\ref{fig8} shows the band ratio maps of NGC\ 2782.  
The intensity ratios of the 6.2 and 7.7\,$\mu$m to the 11.3\,$\mu$m bands
are supposed to be sensitive to the ionization fraction and the size distribution 
of PAHs \citep{draine2001, tielens2008}.
The ratio of S7 to S11 can be regarded as a proxy for this band ratio.
Figure~\ref{fig8}a shows that it is almost constant over the entire galaxy, including the
central and extended regions, at around the values of $\sim 1.2$.
The figure does not indicate any systematic trend with the extended structures.
The constant ratio suggests that the properties of PAHs, in particular their ionization
fraction, do not vary appreciably in the galaxy. 
 
The L24 to S11 color map shows that the ratio is high at the central part of
the galaxy.  This trend is also seen in the spectrum (Figure~\ref{fig4}c).  
The ripple region may have slightly
larger ratios ($\sim 1.8$), but the region with larger ratios
does not show good spatial correlation with the structure seen in the S7 image
(Figure~\ref{fig3}).  The ratio is relatively low at the eastern tail.  
In general, the ratio of the 25\,$\mu$m to 12\,$\mu$m
increases with the incident radiation field and thus with the 
star-formation activity \citep{dale2001, sakon2006, onaka2007a}.  Alternatively, the MIR 
continuum emission also increases with the presence of an AGN \citep{armus2007,
tommasin2010}.
Since no clear sign of AGN is recognized in the spectrum, 
the large L24 to S11 ratio comes more 
likely from the enhanced star-formation in the central part
of the galaxy as suggested by previous investigations \citep[][and references therein]{bravo-guerrero2017}.

Ripples are thought to be formed during galaxy collisions and
several mechanisms have been proposed 
for the ripple formation in mergers.  \citet{smith1994} gives a summary of the mechanisms in
detail, which includes radial oscillations in the
stars and gas due to the head-on collision \citep{wallin1988}, ripples from the debris of the
smaller mass galaxy \citep{quinn1984, hernquist1988}, tidal structures falling back into the remnant \citep{hernquist1992}, stripped matter from a companion disk galaxy \citep{hernquist1988},
a distorted spiral density wave pattern perturbed by an encounter with another galaxy
\citep{thomson1990, howard1993}, and star formation triggered by shocks due to galactic winds 
\citep{fabian1980}. The present result does not show evidence for an increase
in the star-formation in the ripple.

To investigate the spectral energy distribution (SED) of each region in NGC\ 2782 more in detail,
we made photometry at the center, ripple, and tail regions enclosed by the circle and the
solid rectangles in Figure~\ref{fig3}.
The data of N3 are used as a reference point
to estimate the contribution from the stellar component.  
The spectrum of the stellar component is simply assumed to be
given by the average stellar spectrum of the
fluxes of red giant stars used in the flux calibration as in \citet{onaka2010a}.  
The assumed stellar 
contribution is less than 10\% except for the N4 data, in which it amounts to around 40\%.

\begin{figure}[ht!]
\plotone{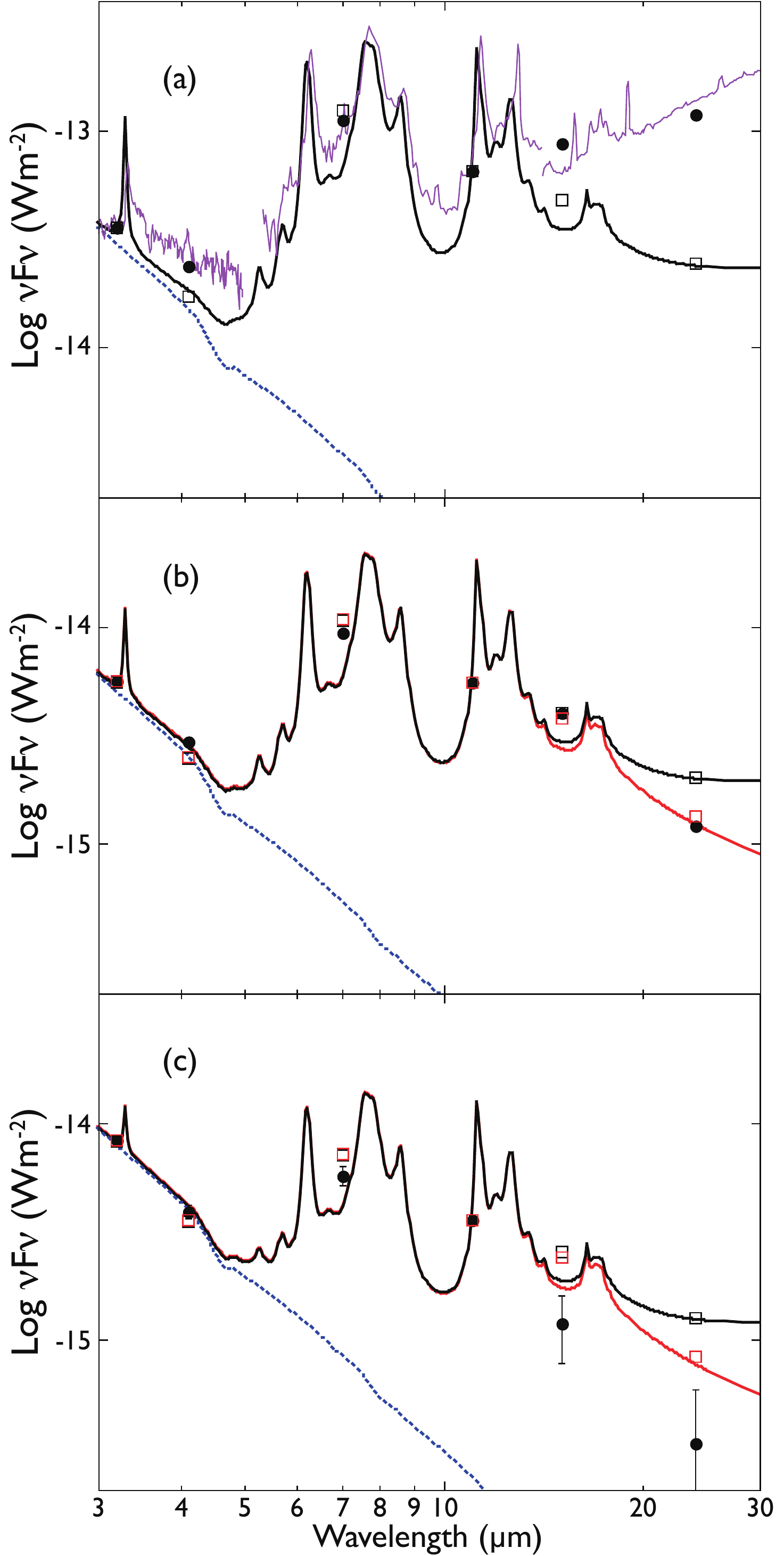}
\caption{SEDs of three regions of NGC\ 2782 indicated in Figure~\ref{fig3}.  (a) Center, (b) ripple, and (c) tail regions.  The results of DUSTEM calculations are shown with the black solid lines for $U=1$ for comparison. They are scaled to the observations at 11\,$\mu$m.  In (a), the IRC and IRS spectra 
of the center region (Figure~\ref{fig4})
are also plotted by the thin purple lines.  The regions where the spectra are extracted are not
exactly the same for the photometry and they are scaled at the photometric data.
In (b) and (c), the red lines are also plotted to show the DUSTEM results
without VSGs (see text). The filled black circles show the IRC photometric results, while
the black and red open squares indicate the color-corrected photometric points of the
DUSTEM calculations corresponding to the black and red lines, respectively, by taking account of 
the IRC filter responses \citep{tanabe2008}.  The blue dotted lines indicate the assumed stellar contribution (see text).
\label{fig9}}
\end{figure}

Figure~\ref{fig9} shows the SEDs of the 
three regions in NGC\ 2782.  As a reference SED we also plot
the results of simple calculations of the DUSTEM model\footnote{\url https://www.ias.u-psud.fr/DUSTEM/} \citep{compiegne2011}.
All the models are scaled to the observations at 11\,$\mu$m.
In the DUSTEM calculation, we assume the standard parameters for the model of the diffuse ISM
of our Galaxy.
We employ the dust properties, size distributions, and ionization fraction of PAHs described 
in \citet{compiegne2011} and assume that the mass fraction
of PAHs (size between 0.35 and 1.2\,nm)
as 8.6\%, and that of very small grains (VSGs, size between 0.6\,nm and 20\,nm) 
as 1.8\%, and that of large grains (LGs, size between 4\,nm and 2\,$\mu$m) as 89.6\%.
We also assume that the incident radiation field strength $U$ in units of the solar neighborhood
is unity.  The SED hardly changes as far as $U$ is less than 10 in this spectral range.
We assume that the flux at N3 is a summation of the dust emission estimated from 
model calculations and the contribution from stellar components
described above.  
In Figure~\ref{fig9}a, the IRC and IRS spectra of the central region (Figure~\ref{fig4}) are
also plotted by the thin purple lines.  For the IRS spectra, the compact and extended components are
added together (\S~\ref{sec:obs}). Since the regions where the spectra are extracted are
not exactly the same as the photometric points, each spectrum is scaled to the 
photometric data points.  There are some differences, but the model spectrum is generally in agreement with the observed spectrum at wavelengths below 13\,$\mu$m, where the spectrum
of models with $U < 10^3$ does not change appreciably.
Excess emission in the NIR relative to the model is also indicated by the
IRC spectrum (see below).
In Figures~\ref{fig9}b and c,
we also plot the results of the model spectrum without VSGs to show the VSG contribution.  
There is no contribution from LGs at wavelengths shorter than 30\,$\mu$m for $U < 10$
\citep{li2001}.

The model flux is slightly larger than the observations at 7\,$\mu$m,
in particular, in the tail region.  It may be attributed to 
a decrease of the ionization degree of PAHs since the emission at
6.2 and 7.7\,$\mu$m is dominated by ionized PAHs, while the emission at 11.3\,$\mu$m comes
mostly from neutral PAHs \citep{draine2001, tielens2008}.  
A low ratio of the PAH 6.2 and 7.7\,$\mu$m to 11.3\,$\mu$m bands are suggested in
halos of galaxies \citep{irwin2006, galliano2008b},
a structure associated with an H$\alpha$ filament in the dwarf galaxy NGC1569 \citep{onaka2010a}, 
and the interarm region of NGC\ 6946 \citep{sakon2007}. The low band ratios
seen in these tenuous regions may have a common origin or arise from
processing of PAHs in these environments.  
The low band ratio in the tail region of NGC\ 2782 is suggested from the medium-band
photometry. 
Spectroscopy of the tail
region is certainly needed to confirm the variation in the band ratio quantitatively.

Figure~\ref{fig9} indicates that
the SED of the center region increases toward longer wavelengths,
while in the ripple and tail regions the SEDs start falling at 15\,$\mu$m.  
The standard diffuse ISM dust model over-predicts the flux at 24\,$\mu$m of the ripple region,
where the model without VSGs better fits the observed SED.  
Comparison with the dust model again supports that the star-formation activity is
not enhanced in the ripple region.

The SED of the tail falls faster than the ripple region
at longer wavelengths and even the model without VSGs
cannot explain the observations at 15 and 24\,$\mu$m well.
Since LGs do not contribute to the emission at 24\,$\mu$m for $U=1$,
it is difficult to explain the observations unless we change the properties of PAHs,
which are the dominant contributor in this spectral region next to VSGs.
Since larger PAHs contribute to emission at longer wavelengths
\citep{draine2007}, we investigate the effect of the size distribution of PAHs in
Appendix~\ref{sect:appendix}.  A simple investigation suggests 
that even extreme models with only very small PAHs
(0.35--0.5\,nm) cannot fully explain the observed SEDs.   The discrepancy at
24\,$\mu$m may be attributed to the model emissivity of PAHs rather than to 
the size distribution.  The emission at 15--24\,$\mu$m 
dominantly arises from stochastically heated VSGs and large PAHs.
Although simple model calculations cannot explain the observed SED of the tail region consistently,
the sharp decline strongly suggests a deficiency of VSGs, and possibly of large PAHs.

In the spectral range 7--30\,$\mu$m, there are several pure rotational transitions of
molecular hydrogen (H$_2$).  The L24 band does not have any H$_2$ lines in its spectral range
(both S(0)\,28.2\,$\mu$m and S(1)\,17.0\,$\mu$m are out of its spectral range),
while L15 includes S(1)\,17.0\,$\mu$m, S11 has S(2)\,12.3\ $\mu$m and S(3)\,9.7\,$\mu$m, and
S7 covers S(4)\,8.0\,$\mu$m, S(5)\,6.9\,$\mu$m, and S(6)\,6.1\,$\mu$m lines.
Therefore, there is a possibility that the H$_2$ line emissions could make
the observed steep decline at 24\,$\mu$m
if they make significant contributions to these bands.
\citet{roussel2007} studied the H$_2$ 0--0 S(0)--S(7) transitions in 57 normal galaxies of
the {\it Spitzer} Infrared Nearby Galaxies Survey \citep[SINGS,][]{kennicutt2003}.
They show that the observed fluxes of the S(0)--S(3) transitions are
in a range (1--400)$\times 10^{-18}$\,Wm$^{-2}$.  They detect the S(4) to S(7) transitions
only in three galaxies in a range (100--250)$\times 10^{-18}$\,Wm$^{-2}$ with
upper limits of about 20$\times 10^{-18}$\,Wm$^{-2}$.
They observe either nuclear regions or bright star-forming complexes 
of the sample galaxies and the observed H$_2$ emissions originate mostly from PDRs.
Larger fluxes ($>50 \times 10^{-18}$\,Wm$^{-2}$) are 
seen toward the galaxies, where star-formation activities are large.
Their observed areas are typically $\sim 300$\,arcsec$^2$, which are about an order of magnitude smaller than the area of the tail region (4200\,arcsec$^2$) to derive the flux for NGC\ 2782. 
The tail region does not have a high star-forming activity (see below).  Thus, we estimate that
the expected fluxes of S(0)--S(3) transitions in the tail region
are at an order of 10$^{-16}$Wm$^{-2}$ at most, even
if we take account of the difference in the size of the observed areas.  
They are 
an order of magnitude smaller than the observed fluxes of the tail region at S7, S11, and L15
(Figure~\ref{fig9}c).
\citet{roussel2007} indicate that the ratio of the summation of the S(0) to S(2)
fluxes to the IRAC band 4 (8\,$\mu$m) flux is relatively constant at around 10$^{-2}$, which
supports the present rough estimate.  Although we cannot
completely rule out the possibility without spectroscopy, 
it does not seem to be very likely that the contribution
from the H$_2$ emission is significant in the tail region.

In the center region, the model with $U=1$ under-predicts
the flux at 4.1\,$\mu$m, which can be attributed to an uncounted component.         
\citet{lu2003} report excess emission in the NIR in normal galaxies, while
similar excess is also seen in the diffuse Galactic emission \citep{flagey2006}.
\citet{smith2009} discuss the origin of excess emission at 4.5\,$\mu$m in dwarf galaxies.
\citet{onaka2010a} investigate excess emission in the NIR of the dwarf galaxy
NGC\ 1569, suggesting that the free-free emission estimated from Br$\alpha$ emission
is insufficient to explain the observed emission and favoring the contribution from hot dust.
Although it probes only a part of the center region of NGC\ 2782, we also estimate the
contribution from the free-free emission using the observed intensity of Br$\alpha$ in the IRC NIR spectrum
(Figure~\ref{fig4}a).  The intensity of Br$\alpha$ is estimated about 
$4.5 \times 10^{-9}$\,W\,m$^{-2}\,$sr$^{-1}$, which suggests the free-free emission being less than 0.05\, MJy sr$^{-1}$.   Therefore, the observed NIR continuum cannot be attributed
to the free-free emission and a contribution from the hot dust component is suggested also for
the center region of NGC\ 2782.                                                                                                                                                                                                                                                                                                                                                                                                                                                                                                                                                                                                                                                                                                                                                                                                                                                                                                                                               

The SED of the center region 
shows excess emission over the model of $U=1$ at 15 and 24\,$\mu$m,
suggesting the presence of warm dust.  
To study the warm dust in the center region quantitatively, we employ data at longer wavelengths.
We retrieved the Post-Basic Calibrated Data of NGC\ 2782
obtained with {\it Spitzer}/MIPS at 70 and 160\,$\mu$m
(AORKEY: 4348416) from the Spitzer Heritage Archive. Since the MIPS 160\,$\mu$m data do 
not have the spatial resolution to resolve the structures of the galaxy, we simply 
made aperture photometry with a radius of $23.\arcsec8$ for all the IRC bands as well as MIPS 70 
and 160\,$\mu$m bands
to match the PSF of the MIPS 160\,$\mu$m.
The sky background is
estimated from the median of the emission outside of the galaxy and subtracted from the
data.  
The total integration time of the MIPS observations
at 70 and 160\,$\mu$m is short, about 120\,s and 42\,s, respectively, and extended emission
is difficult to be detected.  The present aperture photometry is made to derive
a rough estimate of the SED of the
central  $23.\arcsec8$ region of NGC\,2782 from NIR to FIR.

\begin{figure}[ht!]
\plotone{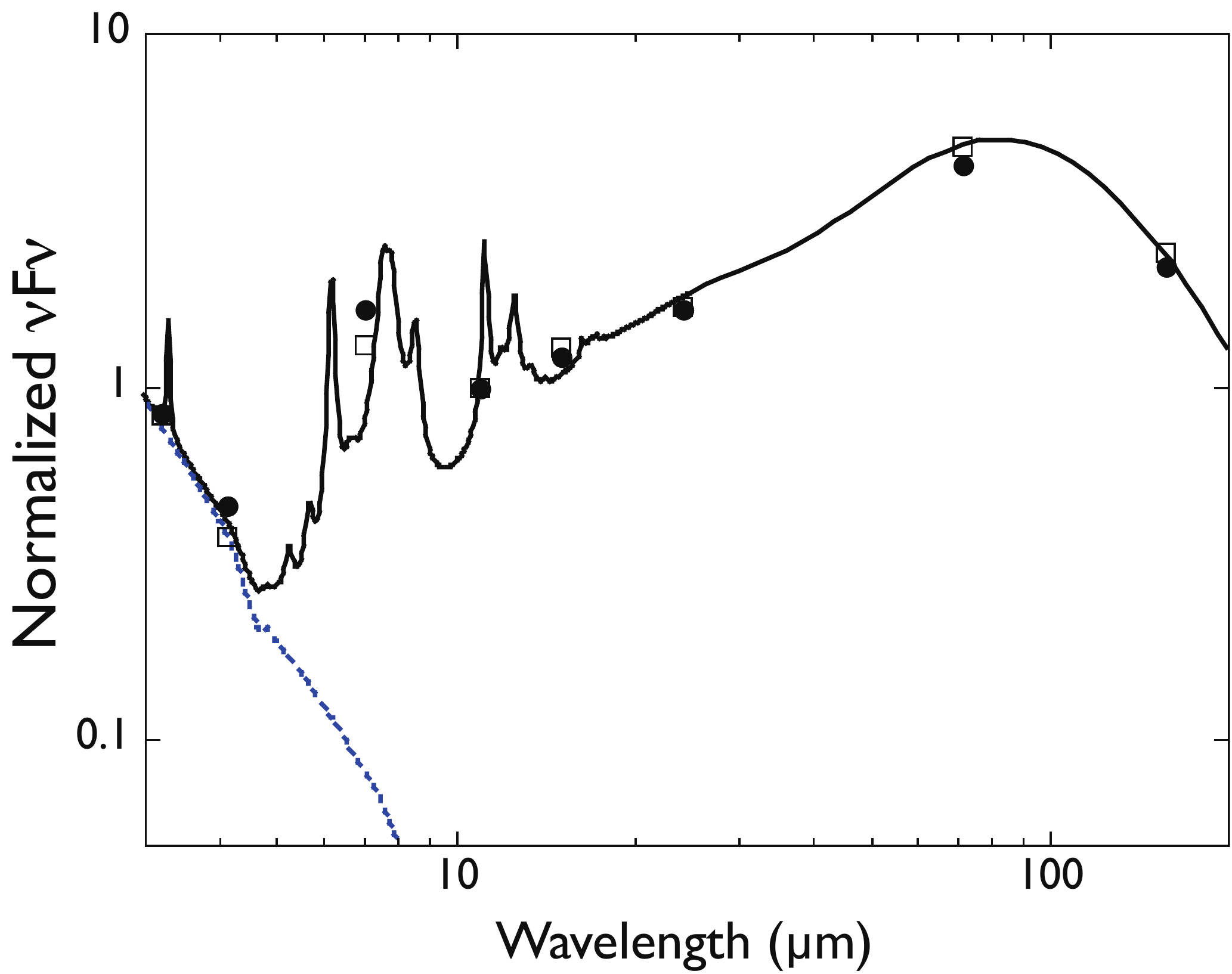}
\caption{SED of the central region of $23$\farcs$8$ in radius (filled circles) from the
IRC and MIPS observations.  
The solid line indicates the result of the DUSTEM calculation (see
text for details).  It is scaled to the observation at 11\,$\mu$m.
The open squares indicate the photometric points of the model spectrum
taking account of the color corrections of the filter responses.  The blue dotted line indicates
the assumed stellar component.
\label{fig10}}
\end{figure}

Figure~\ref{fig10} shows the SED of the center region of the IRC and MIPS observations.  
The SED of the IRC bands is almost identical to that shown in Figure~\ref{fig9}a.
To compare with the DUSTEM model, we
introduce a distribution of $U$ and tune the relative
abundances of PAHs, VSGs, and LGs 
to fit the observation (Figure~\ref{fig10}). The model fits the observations fairly well
if we
assume that the distribution of $U$ is given by a power-law \citep{dale2001}
with the index of $-0.017$,
the minimum and maximum $U$ of 7.6 and $10^6$, respectively, and the mass fractions
of PAHs, VSGs, and LGs are 2.8\%, 8.2\%, and 89\%, respectively.  
The spatial distribution of the infrared emission in galaxies varies with the wavelength and
the available FIR data do not have a spatial resolution to resolve small structures.
Therefore, the derived parameters of the power-law index
and the maximum $U$ should be regarded as rough estimates.  The derived parameters support
an enhanced star-formation activity in the center region.
The derived PAH abundance is small relative to the standard value (8.6\%),
but is still within a range of normal galaxies \citep{draine2007b}.  The small PAH abundance 
might be related to the destruction of PAHs due to the presence of an AGN.  As described above, however, the present IR spectra do not show clear evidence of an AGN.
Only recent X-ray observations indicate the presence of a low-luminosity AGN \citep{tzanavaris2007,
bravo-guerrero2017}.

\citet{knierman2013} detected a number of star cluster candidates (SCCs) 
both in the western and eastern tails.
In the western tail, one H$\alpha$ source is found, but no CO and [\ion{C}{2}]\ 158\,$\mu$m line
emission has been detected.  They attribute the absence of the CO and [\ion{C}{2}] emission
to a low carbon abundance in the western tail.
No detection of the MIR emission in the western tail is compatible with their interpretation. 

In the eastern tail, several H$\alpha$ sources are detected as well as the CO and
[{\ion{C}{2}] emission. \citet{knierman2013} made estimates of 
SFRs of the regions enclosed by the blue dashed rectangles in Figure~\ref{fig3}
from the intensities of H$\alpha$ and [\ion{C}{2}].  Here we use the PAH emission to estimate SFRs
of these regions.  We measure the fluxes in the S7 band for each region
(TDGC, HI-N, HI-M, and HI-S, see Figure~\ref{fig3}) and subtract the assumed stellar contribution.  
The stellar contribution is at most 6\% and thus does not affect the conclusion.
We assume that the emission
at S7 mainly comes from the PAH 6.2\,$\mu$m and 7.7\,$\mu$m bands and apply the PAH band-SFR
relations provided by \citet{shipley2016} to estimate the SFR, assuming that the distance is 39.5\,Mpc
\citep{knierman2013}.  The S7 band collects almost 
the entire PAH 6.2 and 7.7\,$\mu$m band fluxes, but it could also have a contribution from
other components.  Thus, the SFR from the S7 band should be taken as an upper limit.
Also note that the spatial resolution of the present observation is different from those of
 \citet{knierman2013} and the regions where the estimate made
are not exactly the same.  

\begin{figure}[ht!]
\plotone{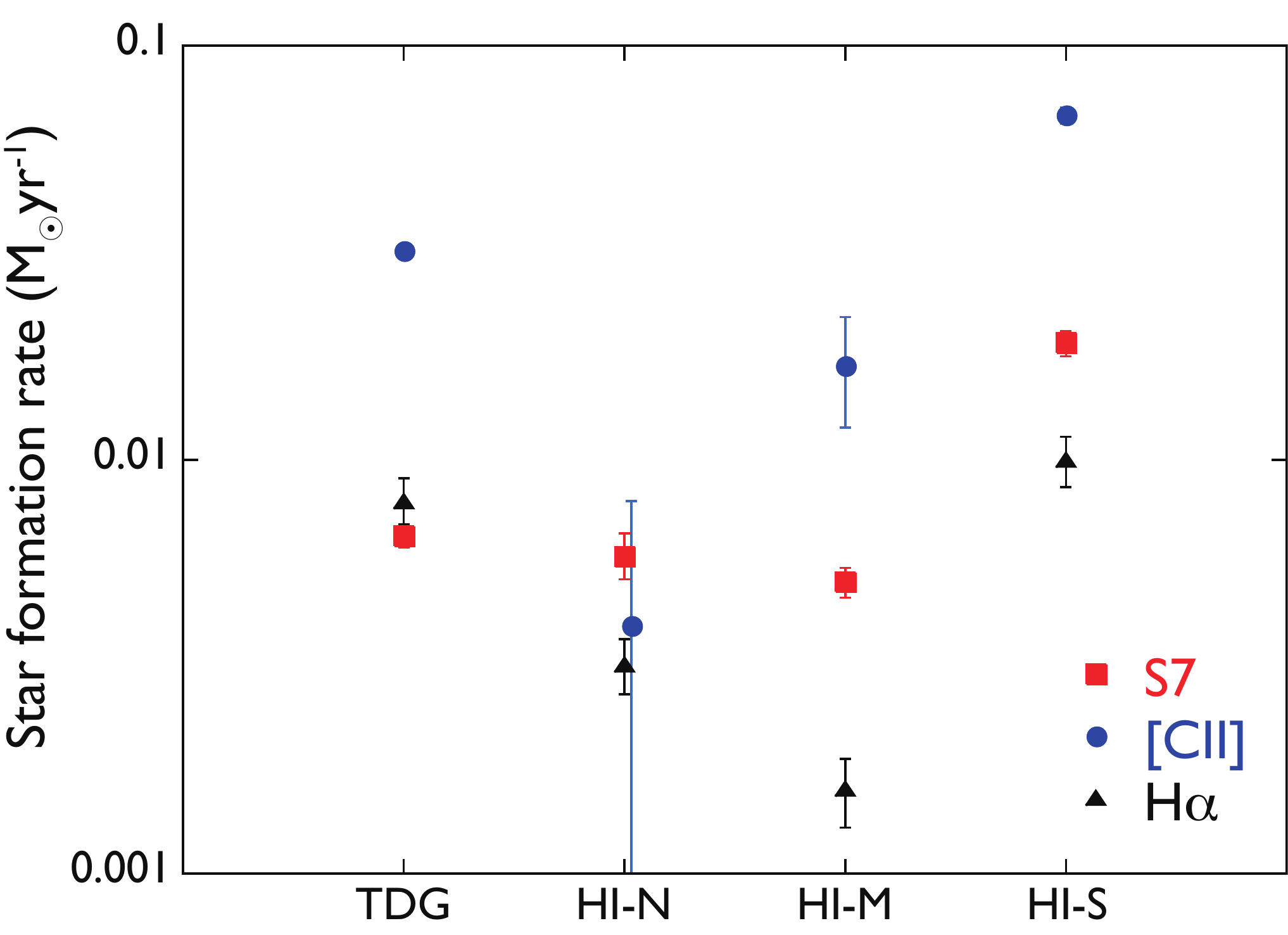}
\caption{Star-formation rates for the regions indicated in Figure~\ref{fig3}.  The red squares
show those estimated from the PAH bands of the 
present S7 image (see text for details), while the black triangles and the blue circles indicate those from H$\alpha$ and [\ion{C}{2}], respectively \citep{knierman2013}.
\label{fig11}}
\end{figure}

Figure~\ref{fig11} shows
comparison of the SFRs thus estimated from the flux at S7 with those from
H$\alpha$ and [\ion{C}{2}] for each region.  It indicates that the SFR from the PAH bands (S7)
is close to that from H$\alpha$ (TDGC and HI-N) or in between those from H$\alpha$ and [\ion{C}{2}]
(HI-M and HI-S).  A good correlation between the PAH emission
and [\ion{C}{2}] has been reported for star-forming galaxies, which can be interpreted
if the PAHs are the dominant agency for gas heating \citep{helou2001}.  The H$\alpha$ 
emission indicates the strength of the heating radiation.  
Taking account of the uncertainties described above, the agreement is
sufficiently good.  The present results suggest that 
the PAH emission may be used as a useful measure also for the
SFR in these extended structures, although a change in the dust size distribution is
suggested.  However, the present sample size is too small to draw a general conclusion.
We definitely need much more samples to confirm its applicability.
To investigate the size distribution and its effect on the estimate of the SFR
quantitatively, we need FIR
data with a sufficient spatial resolution and sensitivity that allow us to resolve and detect
emission from extended structures in the FIR. 

\subsection{NGC\ 7727} \label{sec:discussion:NGC7727}
Figure~\ref{fig12} shows S7 to S11 and L24 to S11 color maps of NGC\ 7727. 
The S7 to S11 color is almost constant over the entire galaxy.  The L24 to S11 color does not
change over the galaxy except for a slight increase in a very small region around the center.
The plume regions in the north may suggest a small increase in the band ratio.
Compared to the color maps of NGC\ 2782 (Figure~\ref{fig8}), the L24 to S11 band ratio 
of NGC\ 7727 is
small, suggesting low star-forming activity in NGC\ 7727.

\begin{figure*}[ht!]
\plotone{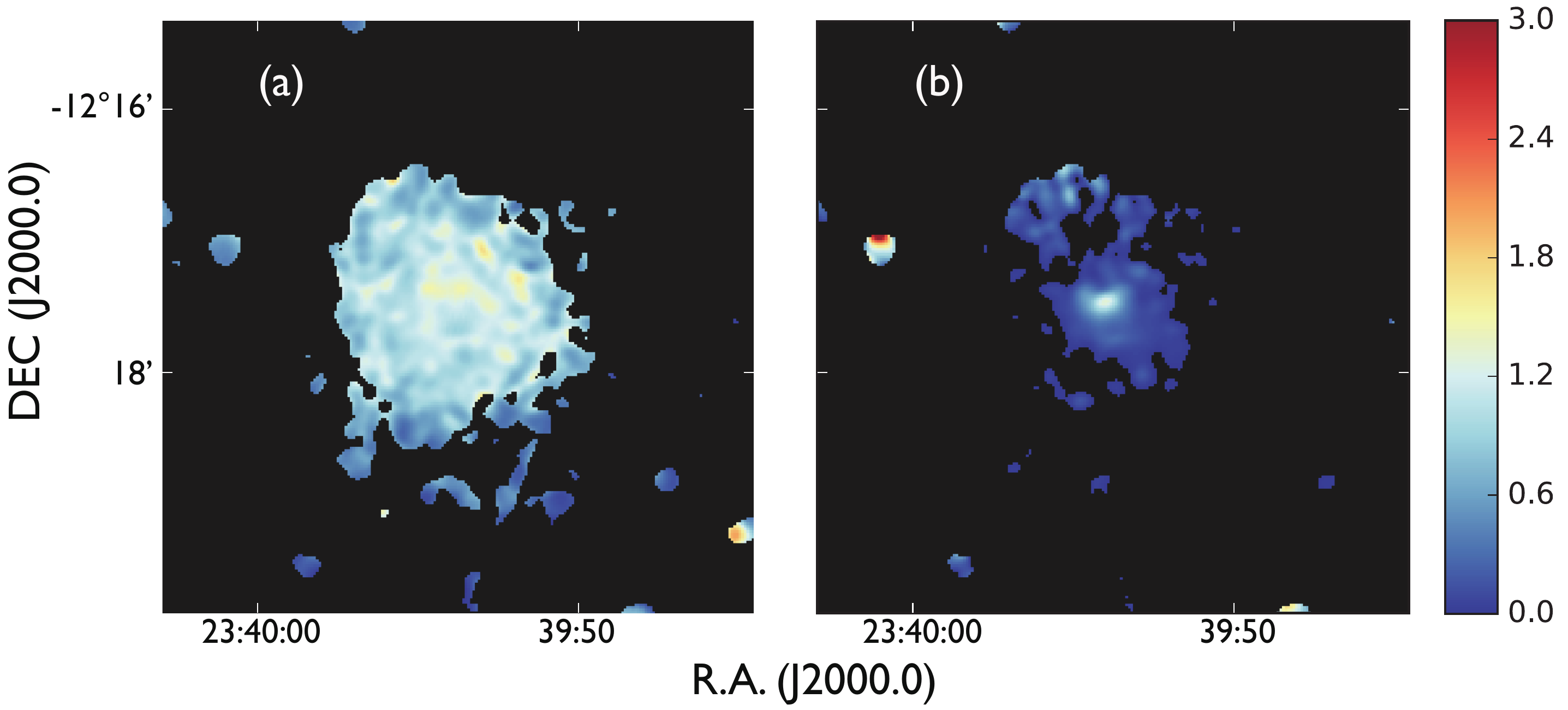}
\caption{Band ratio maps of NGC\ 7727. (a) S7 to S11 and (b) L24 to S11.  
The pixels with fluxes less than 4$\sigma$ are masked and shown in black.
\label{fig12}}
\end{figure*}

\begin{figure}[ht!]
\plotone{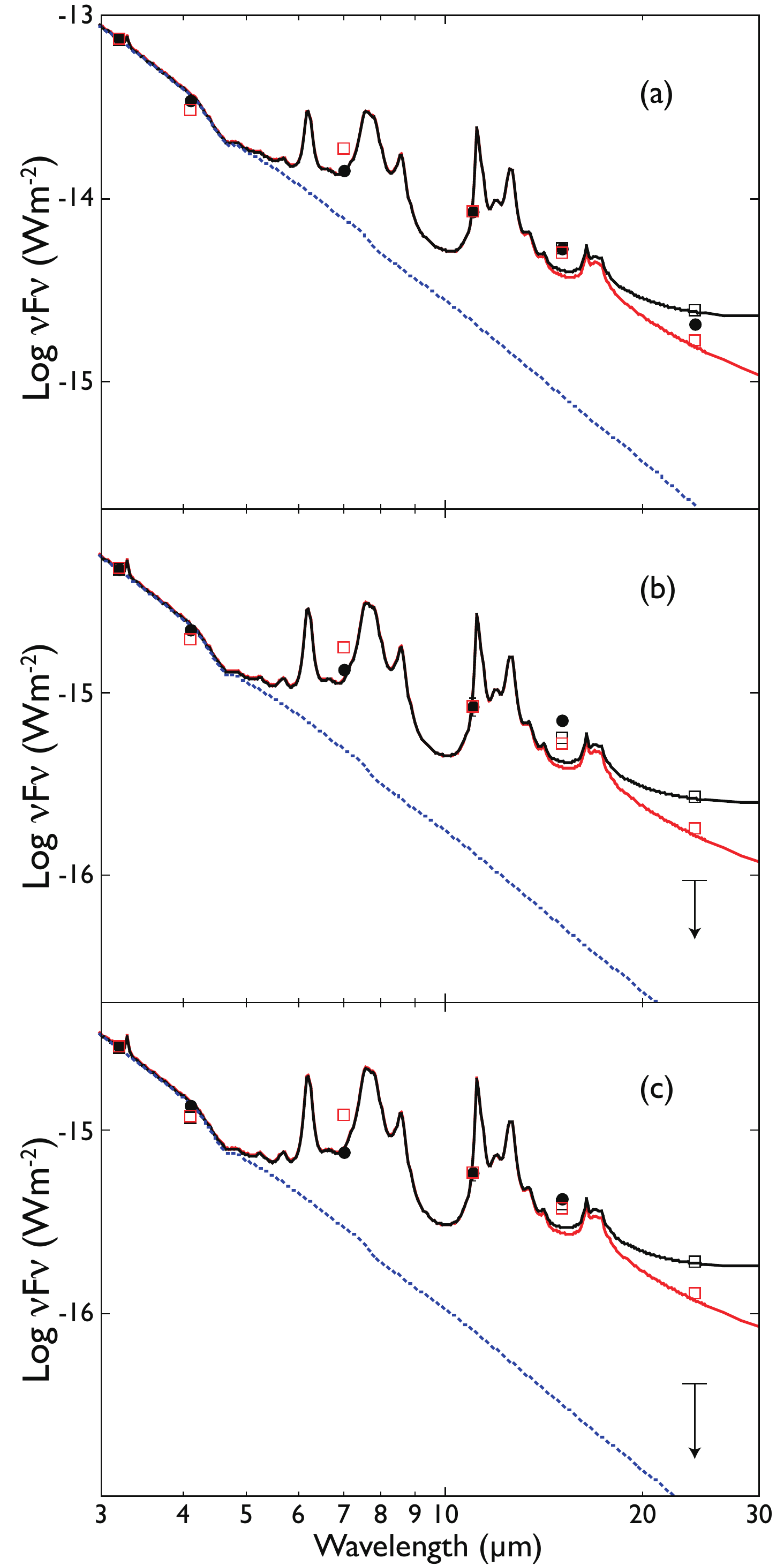}
\caption{SEDs of three regions of NGC\ 7727 indicated in Figure~\ref{fig6}.  (a) Center, (b) plume 1, and (c) plume 2 regions.  The results of DUSTEM calculations are shown by 
the black and red lines for $U=1$ for the models with and without VSGs, respectively. 
They are scaled to the observations at 11\,$\mu$m.
The filled black circles show the IRC photometric results, while
the black and red open squares indicate the color-corrected photometric points of the
DUSTEM calculations corresponding to the black and red lines, respectively.  
The blue dotted lines indicate the assumed stellar contribution.
\label{fig13}}
\end{figure}

Figure~\ref{fig13} shows the SEDs of the three regions indicated by the circle and the
solid rectangles in Figure~\ref{fig6}.  
The same spectra of the DUSTEM model as in Figure~\ref{fig9} are also plotted with the
assumed stellar contribution.  
Compared to NGC\ 2782, the contribution from the stellar continuum is estimated to be large.
The data at N4 (4.1\,$\mu$m) are well explained by the model,
suggesting that the assumed spectrum well approximates the stellar contribution.
The NIR spectrum of the center shows a steeply decreasing continuum toward longer wavelengths
(Figure~\ref{fig7}).  This is in agreement with the dominance of the
stellar contribution.  The DUSTEM model predicts weak PAH emission at 3.3\,$\mu$m
(Figure~\ref{fig13}a}), which is not seen in the spectrum.  This may suggest the absence of small PAHs that predominantly
emit the 3.3\,$\mu$m band.

The contribution from the H$_2$ emission can be estimated in a manner similar to \S\ref{sec:discussion:NGC2782}.  Since the sizes of the areas from which the fluxes are
estimated in the plume regions of NGC\ 7727 are about the same (600 and 324\,arcsec$^2$, 
for plume 1 and plume 2, respectively) as those in
the observations of the SINGS galaxy sample \citep{roussel2007}, we expect that the contribution
from the H$_2$ emission
is about an order of magnitude smaller (an order of $10^{-17}$\,Wm$^{-2}$)
than the observed fluxes (Figures~\ref{fig13}b and c) unless
the plume regions have high star-formation activities.  Spectroscopic studies are 
certainly needed for the confirmation.

There are clear discrepancies in the flux at 7\,$\mu$m for all the three regions, which are larger
than that in the tail region of NGC\ 2782.
The model (indicated
by open squares) over-predicts the flux at 7\,$\mu$m relative to that at 11\,$\mu$m.
Since 
the observed S7 to S11 ratio seems to be almost the same as NGC\ 2782 (Figures~\ref{fig8}a and
\ref{fig12}a),
the difference should be attributed to the larger contribution of the stellar component
in NGC\ 7727.  The stellar spectrum has a steep spectral dependence and thus
the ISM contribution of NGC\ 7727 becomes flatter at 7--11\,$\mu$m compared to NGC\ 2782.
The resultant ISM SED cannot be explained by the standard PAH emission spectrum.  
The relatively constant S7 to S11 color suggests that this characteristic of the SED may hold 
for the entire galaxy.

In addition to low band ratios seen in halo regions, a filament, and an interarm region, 
elliptical galaxies show peculiar MIR spectra that have
the clear PAH 11.3\,$\mu$m emission with very weak or almost no emission 
at 6.2, 7.7, and 8.6\,$\mu$m,
which are the strongest bands in normal PAH emission spectra
\citep{kaneda2005, kaneda2007}.
The IRC NIR spectroscopy also shows that 
the 3.3\,$\mu$m is absent in the elliptical galaxy NGC1316 \citep{kaneda2007}.  
These characteristics may be interpreted in terms of emission
from large and neutral PAHs.  
These characteristics seem to accord with the observed MIR SEDs of NGC\ 7727.
In NGC\ 7727, large and neutral PAHs may dominate
due to the weak star-formation activity. 
NGC\ 7727 is classified as SAB(s)a pec \citep{RC3}, and it may be in a stage to 
an elliptical galaxy after merging.  
We certainly need spectroscopy to confirm this interpretation.

The emission of the center region at 11--24\,$\mu$m can be explained by the DUSTEM
model with $U=1$, while the SED of the plume regions declines very sharply at 
24\,$\mu$m compared to the DUSTEM model.
The fluxes at 24\,$\mu$m for the both plumes are less than 3$\sigma$ and upper limits
are plotted in Figures~\ref{fig13}b and c.  If the fluxes are at the levels that the models predict,
they should be detectable and thus the sharp decline at 24\,$\mu$m is secure.
This is a similar trend seen in the tail region of NGC\ 2782.  Figures~\ref{fig13}b and c 
suggest that even models without VSGs cannot explain the observed SED at 24\,$\mu$m.
We discuss the effects of the size distribution of PAHs in Appendix~\ref{sect:appendix}
similarly to NGC\ 2782.  It seems difficult to explain the observed SEDs solely by the change
of the size distribution and the revision of the PAH emissivity at 15--24\,$\mu$m may be needed.
Although it may not be typical PAH emission, excess emission at 7--15\,$\mu$m 
is certainly present (Figure~\ref{fig13}).  The sharp decline at 24\,$\mu$m relative
to the excess suggests that the dust that emits predominantly at 24\,$\mu$m is deficient
in the plume regions of NGC\ 7727.

\subsection{Formation and destruction of PAHs and VSGs}\label{sec:discussion:PAH}
In both galaxies, excess emission at 7--15\,$\mu$m and a sharp decline of the emission at 24\,$\mu$m are seen in the extended structures.  
The sharp decline in the extended structures relative to the Milky Way and the center regions of the galaxy is secure based on the present observations, 
though the contribution from line emission needs to be investigated by 
spectroscopy.  If the observed SEDs originate mostly from the dust emission that the DUSTEM 
model simulates, then the decline must be interpreted in terms of the variation in the
dust size distribution.
The absolute abundance of VSGs
depends on the assumed emissivity, but comparison with the DUSTEM model indicates that
these regions have a small abundance of VSGs relative to 
the diffuse ISM of our Galaxy and the center regions of the galaxy.  Since those extended structures are thought to have been
formed in
the merger events, the deficiency of VSGs must be related to the events.
The excess emission at 7--15\,$\mu$m is attributed to the PAH features. 
In general VSGs are
assumed to consist of amorphous carbonaceous dust \citep{compiegne2011}.
The presence of PAHs and the deficiency of VSGs suggest a possible scenario that 
VSGs fragmented into PAHs in the structures formed during the merger events.
The formation of PAHs from fragmentation of carbonaceous dust has been
discussed theoretically \citep{jones2013, seok2014} and suggested in various celestial objects
observationally \citep{onaka2010a, seok2012, lau2016}.  \citet{pilleri2012, pilleri2014} also study 
the photo-fragmentation of VSGs into PAHs in PDRs.
FIR data with a sufficient spatial resolution and sensitivity are needed to obtain a clear answer if the fraction of PAHs increases relative to LGs by fragmentation and/or if the entire dust size distribution is changed (\S\ref{sec:discussion:NGC2782}). 

If the observed SEDs are the results of fragmentation of
VSGs into PAHs during the merger events, the size distribution of PAHs and VSGs
must have been preserved since then.  The proposed scenario
requires that the destruction time scale of PAHs and the formation time scale
of VSGs should be longer than the merger ages.
The merger age is estimated as about 200--300\,Myr 
for NGC\ 2782 \citep{smith1994, knierman2013} and 1.3\,Gyr for NGC\ 7727
\citep{georgakakis2000}.  In our Galaxy, the lifetime of carbonaceous dust is estimated as 170\,Myr
\citep{serra diaz-cano2008} and that of PAHs is as $\sim 100-500$\,Myr
\citep{micelotta2010a}.  Recently \citet{bocchio2014} re-evaluate that the lifetime of carbonaceous
dust is even shorter as $\approx 62\pm 56$\,Myr by merging the inertial and thermal sputtering
into a single process.  We try a very rough estimate to translate the lifetime in our Galaxy
to the tail region in NGC\ 2782 and the plume regions in NGC\ 7727.
Supernova shocks are the dominant destruction process and thus
the lifetime must be related to the SFR. 
The estimated total SFRs in the tail region of NGC\ 2782 is 0.02--0.13
\,M$_\odot$\,yr$^{-1}$ \citep[Figure~\ref{fig11},][]{knierman2013}.
\citet{chomiuk2011} collect various estimates of the SFR of our Galaxy and derive an average
value of $1.9 \pm 0.4$\,M$_\odot$yr$^{-1}$, while \citet{licquia2015} apply a hierarchical
Bayesian statistical method and obtain $1.65 \pm 0.19$\,M$_\odot$yr$^{-1}$.
The area of the tail region indicated in Figure~\ref{fig3} is about 150\,kpc$^2$ at a distance
of 39.5\,Mpc, which is about half a size of the surface area of our Galaxy disk.
If it is inversely proportional to an average SFR per area, the destruction time scale
in the tail region can be by an order of magnitude longer than the Galactic value.
Thus, it may be longer than the merger age.
We do not have much information on the SFR in the plume regions of NGC\ 7727.  
Although their PAH features do not seem to be typical and thus it has a large uncertainty, 
we use the S7 band flux to infer their SFRs.
The observed fluxes at S7 in the plume regions are smaller by about an order of magnitude
than that of the tail region of NGC\ 2782 (Figures~\ref{fig9}c, \ref{fig13}b, and c), 
suggesting that the SFR per area is
also smaller.  
The lifetime of PAHs in these regions may be two orders of magnitude longer than
the Galactic value and thus may be longer than its merger age.

The estimated very short destruction time scale of carbonaceous 
dust in our Galaxy 
suggests rapid recycling of carbonaceous dust in our Galaxy.
To compensate the rapid destruction, it is generally
suggested that dust grows in dense clouds in a similar timescale \cite[e.g.,][]{jones2011, bocchio2014}.
Direct conversion of these values into the formation
time scales of carbonaceous dust in the tail and plume regions is not 
straightforward, but these investigations suggest that 
formation of carbonaceous dust, thus VSGs, 
requires dense clouds and must also be related to the
star-formation activity.  The suggested low SNRs per area in the extended regions of both galaxies
indicate a longer time scale also for the formation of carbonaceous dust, including VSGs.
Although these estimates are very rough, 
the destruction of PAHs and the formation
of VSGs may take a longer time than the merger ages 
due to the low star-formation activities for both galaxies.  Then,
the size distribution of PAHs and VSGs may have been preserved after the
merger events until now.

It should be noted that the above discussion is made on the basis of medium-band
photometry.  While emission of molecular hydrogen should not contribute to the present
results significantly as discussed above, contributions from other lines or features cannot
be ruled out completely.  Spectroscopic study is definitely needed to support the above
discussion and interpretation.

\section{Summary}
We present NIR to MIR imaging and NIR spectroscopic observations of the two mergers,
NGC\ 2782 and NGC\ 7727, with {\it AKARI}/IRC.  
The MIR to FIR SED suggests enhanced star-formation activities in the central part of NGC\ 2782.
No evidence for the presence of an AGN is obtained in the present observations.
The ripples seen in optical images of NGC\ 2782, which are thought to have been 
formed by the merger
event, are also seen in the MIR images.  The SED suggests no enhancement in the star-formation
in the ripples.  The IR SED of NGC\ 7727 is dominated by the stellar component.  Neither PAH 3.3\,$\mu$m emission nor \ion{H}{1} recombination lines are detected in the central part of the galaxy,
suggesting a low star-formation activity.

NGC\ 2782 shows extended emission
(tail structure)
at 7--15\,$\mu$m, whose structure well corresponds to the eastern tidal tail seen in the
\ion{H}{1} map.  NGC\ 7727 shows extended emission (plumes) at 7--15\,$\mu$m, similar structure to which 
has been seen in the $K$-band image.  Both structures are thought to have been
formed by the merger events.  The SED of the tail of NGC\ 2782 at 7--15\,$\mu$m can well be
explained by PAH emission with
a smaller ionization degree of PAHs relative to the diffuse ISM of our Galaxy.
The SEDs of the plume regions of NGC\ 7727 suggest PAH emission of much lower ionization degree.
It may be similar to those seen in 
elliptical galaxies.  Spectroscopic studies are definitely needed to confirm the nature of
their emission.

Extended emissions in both galaxies decline more rapidly
at 24\,$\mu$m than the model SED of the ISM of our Galaxy,  
suggesting a deficiency of VSGs.  
While it must be confirmed by spectroscopy that
emission lines or other features do not contribute to the observed fluxes significantly, 
the observed SEDs of the extended
emission can be interpreted if VSGs fragment into PAHs 
during the merger events. Because of the low star-formation activities in both
regions, the time scales for the destruction of PAHs and formation of VSGs may be
longer than the merger ages.  Then, the present infrared SEDs
retain the imprint of the size distribution of PAHs and
VSGs created during the merger events.

The star-formation rate of the tail region of NGC\ 2782
estimated from the 7\,$\mu$m band is in agreement with those
estimated from H$\alpha$ and [\ion{C}{2}]\ 158\,$\mu$m.  The present results suggest
that the PAH emission may also be used as
a star-formation rate measure for extended structures of galaxies despite possible
processing of dust grains in these regions.  However, the number of the present sample is
too small to draw a general conclusion.  We certainly need much larger samples 
to test its applicability.
The present results also suggest that MIR observations are very efficient
to identify and study extended structures formed by merger events and to
study processing of dust in merger events.  Further studies with FIR data that allow
us to quantitatively investigate
the size distribution of dust in structures associated with merger events
will provide a unique opportunity to investigate the dust lifetime and processing
in a galactic time scale and its effect on the estimate of the SFR from the PAH emission.

\acknowledgements
This work is based on observations with {\it AKARI}, a JAXA project
with the participation of ESA.  The authors thank all the members of the {\it AKARI} project
for their continuous support.  
This research has made use of the NASA/IPAC Infrared Science Archive and the NASA/IPAC 
Extragalactic Database (NED), both of which are operated by the Jet Propulsion Laboratory, California Institute of Technology, under contract with the National Aeronautics and Space Administration.
R. W. was supported by the Japan Society for the Promotion of Science (JSPS) as an
International Research Fellow.
This work is supported by JSPS and CNRS under the Japan -- France Research
Cooperative Program and a grant
for Scientific Research from the Japan Society for the Promotion of Science
(nos. 23244021).

\vspace{5mm}
\facilities{AKARI, Spitzer}

\appendix
\section{Model calculations of modified size distribution of PAH}\label{sect:appendix}
\begin{figure*}[ht!]
\plotone{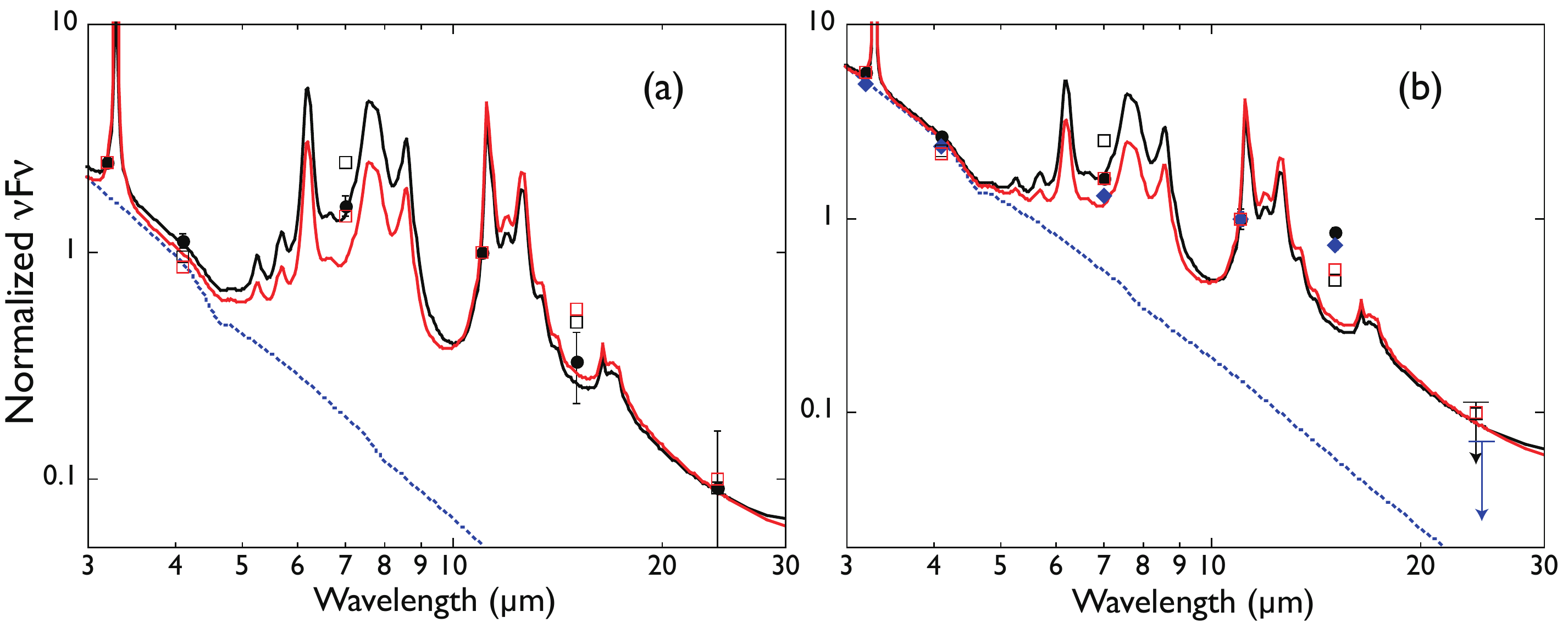}
\caption{Normalized SEDs of (a) the tail region of NGC\ 2782 (filled circles) 
and (b) the plume regions of NGC\ 7727 (the filled black circles for plume 1 and
the filled blue diamonds for plume 2).
The black and red lines show the DUSTEM model with only small PAHs and
that with small, neutral PAHs, respectively (see text). 
The black and red open squares indicate the photometric points of the model spectra of
the black and red lines, respectively, 
taking account of the color corrections of the filter responses.  The blue dotted lines indicate
the assumed stellar component.  The spectra are normalized at 11\,$\mu$m for comparison.
\label{figA1}}
\end{figure*}

To explain the sharp decline seen at 15--24\,$\mu$m in the SEDs of the tail region of NGC\ 2782 and
the plume regions of NGC\ 7727, we investigate the effects of the PAH size distribution
by a simple model analysis.  
The size distribution of PAHs in the original model
is given by a log-normal distribution between 0.35\,nm and 1.2\,nm.  
Larger PAHs contribute to emission at longer wavelengths
\citep{draine2007}.  We decrease the maximum size and compare the model
with the observations in Figure~\ref{figA1} (black lines).  The spectra are normalized at 11\,$\mu$m for comparison.  
We found that the observed flux at 24\,$\mu$m relative to that at 11\,$\mu$m
can be explained if we assume the presence of 
only small PAHs of 0.35--0.5\,nm.  
The assumed size distribution is very extreme and contrived, and is used only for investigation of 
the size effect.
However, even the model that has such an extreme size distribution
cannot be reconciled with the observed flux at 15\,$\mu$m satisfactorily.
In addition, because of the size distribution biased toward smaller PAHs, the model flux
at 7\,$\mu$m exceeds well above the observation.

The 6.2 and 7.7\,$\mu$m PAH
bands are enhanced when PAHs are ionized \citep{draine2001, tielens2008}.  In the standard model, the size-dependent
ionization fraction of PAHs is assumed and the ionization fraction of small PAHs is
around 38\% \citep{compiegne2011}.
To investigate the effect of the ionization fraction in addition to the size distribution, 
we also calculate the model with the same
size distribution of only small PAHs, assuming that they are all neutral.  The resultant model
spectrum is shown by the red lines in Figure~\ref{figA1}.  Small neutral PAH model
may explain the observations at 7, 11, and 24\,$\mu$m, but the discrepancy
at 15\,$\mu$m is increased.  
It should also be noted that the models with only small PAHs predict 
very strong PAH emission at 3.3\,$\mu$m, which should be detectable in the N3 image.
Taking account of these facts, simple modifications of the size distribution
and the ionization degree of PAHs have difficulties in explaining the observed SEDs consistently,
suggesting that the modification of the PAH emissivity
may be needed to fully explain the observed SEDs of the tail of NGC\ 2782 and
the plume regions of NGC\ 7727.

\listofchanges

\end{document}